%% file: main_icml.tex
\documentclass{article}

\usepackage{microtype}
\usepackage{graphicx}
\usepackage{subfigure}
\usepackage{booktabs} 

\usepackage{hyperref}

\usepackage[accepted]{icml2024}

\usepackage{amsmath}
\usepackage{amssymb}
\usepackage{mathtools}
\usepackage{amsthm}

\usepackage{colortbl}
\usepackage{xcolor}
\usepackage{multicol}
\usepackage{multirow}
\input{math_commands}

\usepackage{tablefootnote}
\usepackage{booktabs}
\usepackage{makecell}

\usepackage[capitalize,noabbrev]{cleveref}

\theoremstyle{plain}
\newtheorem{theorem}{Theorem}[section]
\newtheorem{proposition}[theorem]{Proposition}
\newtheorem{lemma}[theorem]{Lemma}

\theoremstyle{definition}
\newtheorem{definition}[theorem]{Definition}

\theoremstyle{remark}

\usepackage[textsize=tiny]{todonotes}

\newcommand{\model}{EquiCSP}

\icmltitlerunning{Equivariant Diffusion for Crystal Structure Prediction}

\begin{document}

\twocolumn[
\icmltitle{Equivariant Diffusion for Crystal Structure Prediction}



\icmlsetsymbol{equal}{*}

\begin{icmlauthorlist}
\icmlauthor{Peijia Lin}{sysu}
\icmlauthor{Pin Chen}{sysu,nscc}
\icmlauthor{Rui Jiao}{TS,AIR}
\icmlauthor{Qing Mo}{nscc}
\icmlauthor{Jianhuan Cen}{sysu}
\icmlauthor{Wenbing Huang}{renmin,BeijingLab}
\icmlauthor{Yang Liu}{TS,AIR}
\icmlauthor{Dan Huang}{sysu}
\icmlauthor{Yutong Lu}{sysu,nscc}
\end{icmlauthorlist}

\icmlaffiliation{sysu}{School of Computer Science and Engineering, Sun Yat-sen University, Guangzhou, China}
\icmlaffiliation{nscc}{National Supercomputer Center in Guangzhou, China}
\icmlaffiliation{TS}{Dept. of Comp. Sci. \& Tech., Institute for AI, Tsinghua University, Beijing, China}
\icmlaffiliation{AIR}{Institute for AIR, Tsinghua University, Beijing, China}
\icmlaffiliation{renmin}{Gaoling School of Artificial Intelligence, Renmin University of China, Beijing, China}
\icmlaffiliation{BeijingLab}{Beijing Key Laboratory of Big Data Management and Analysis Methods, Beijing, China}


\icmlcorrespondingauthor{Yutong Lu}{luyutong@mail.sysu.edu.cn}

\icmlkeywords{Crystal Structure Prediction, Equivariant Graph Neural Networks, Diffusion Model, AI for Science}

\vskip 0.3in
]



\printAffiliationsAndNotice{}

\begin{abstract}
 In addressing the challenge of Crystal Structure Prediction (CSP), symmetry-aware deep learning models, particularly diffusion models, have been extensively studied, which treat CSP as a conditional generation task. However, ensuring permutation, rotation, and periodic translation equivariance during diffusion process remains incompletely addressed. In this work, we propose \model{}, a novel equivariant diffusion-based generative model. We not only address the overlooked issue of lattice permutation equivariance in existing models, but also develop a unique noising algorithm that rigorously maintains periodic translation equivariance throughout both training and inference processes. Our experiments indicate that \model{} significantly surpasses existing models in terms of generating accurate structures and demonstrates faster convergence during the training process. Code is available at \url{https://github.com/EmperorJia/EquiCSP}.
\end{abstract}

\section{Introduction}
Crystal structure prediction (CSP) seeks the atomic arrangement with the lowest energy for given chemical compositions and conditions~\citep{desiraju2002cryptic}, focusing on finding the global minimum of the potential energy surface. This task, while conceptually straightforward, is a significant challenge in physics, chemistry, and materials science due to the complexity of the potential energy landscape and the exponential increase in possible structures with more atoms in a unit cell~\cite{oganov2019structure}.

Traditional CSP methods primarily employ Density Functional Theory (DFT)~\citep{kohn1965self} for iterative energy calculations, integrating optimization algorithms like genetic algorithm~\cite{2006Crystal,2011How} and particle swarm optimization~\cite{2010Crystal,WANG20122063} to navigate the energy landscape for stable states. However, the time-intensive nature of DFT calculations makes these traditional CSP approaches notably inefficient.

Recent advances have seen a shift towards deep generative models, which learn distributions directly from datasets of stable structures~\cite{court20203,yang2021crystal}, with diffusion models, a subset of these models, gaining prominence in crystal generation~\cite{xie2021crystal,jiao2023crystal}. 
Diffusion models are lauded for their superior interpretability and performance, owing to their inherent physical explainability.
However, developing diffusion models for CSP involves addressing specific challenges. Physically, E(3) transformations, including translation, rotation, and reflection of crystal coordinates, do not change physical laws, necessitating E(3) invariant sample generation in the model. 
Typical diffusion models like denoising diffusion probabilistic models (DDPMs)~\citep{sohl2015deep,vignac2023digress} and score-based generative models with stochastic differential equations (SDEs)~\citep{song2020score}, initially used in computer vision, require E(3) equivariance when adapted to molecular graph domains~\citep{luo2021predicting}, and for crystals, additional consideration of periodic invariance is needed~\citep{jiao2023crystal}.
In this study, we introduce \model{}, an equivariant\footnote{This paper mainly focuses on lattice permutation equivariance and periodic translation equivariance, which are unique to crystallographic data. Traditional rotation equivariance is addressed using invariant representations.} diffusion method to address CSP. 
We track the impact of lattice permutation in crystals on diffusion models used for CSP and propose corresponding solutions.
\model{} entails that during generation and training, any permutation of lattice parameters results in a corresponding equivariant transformation of the atomic fractional coordinates. Furthermore, we propose a specialized diffusion noising algorithm in SDEs, meticulously designed to preserve periodic translation equivariance consistently during both training and inference stages.

To summarize, our contributions in this work are as follows:
\begin{enumerate}
    \item To our knowledge, we are the first to address lattice permutation equivariance in diffusion models, aiming to address that any permutation of lattice parameters during training and generation corresponds to an equivalent transformation in atomic fractional coordinates, achieved through simple and efficient loss functions rather than encoding the equivariance directly into the neural networks.
    \item In addition, we propose a novel diffusion noising method, named Periodic CoM-free Noising, to make the widely used Score-Matching method in SDEs achieving the periodic translation equivariance of crystal generation.
    \item We validate \model{}'s effectiveness in CSP tasks, demonstrating its superior performance over existing learning methods (e.g. CDVAE~\citep{xie2021crystal} and DiffCSP~\citep{,jiao2023crystal}). Furthermore, we enhance \model{} for ab initio generation, proving its enhanced performance comparing to similar methods.
\end{enumerate}

\section{Related Works}

\textbf{Crystal Structure Prediction.} 
Traditional computational methods, such as DFT combined with optimization algorithms, are used to search for local minima on the potential energy surface~\citep{pickard2011ab, yamashita2018crystal, wang2010crystal, zhang2017computer}. Despite their accuracy, these methods are computationally demanding. Recently, machine learning has emerged as an alternative, using crystal databases to predict energy more efficiently than DFT~\citep{jacobsen2018fly, podryabinkin2019accelerating, cheng2022crystal}. Another approach employs deep generative models to directly learn stable structures, representing crystals with 3D voxels~\citep{court20203, hoffmann2019data, noh2019inverse}, distance matrices~\citep{yang2021crystal, hu2020distance, hu2021contact}, or 3D coordinates~\citep{nouira2018crystalgan, kim2020generative, REN2021}. However, these methods often overlook the complete symmetries in crystal structures.

\textbf{Equivariant Graph Neural Networks.} 
E(3) symmetric, geometrically equivariant Graph Neural Networks (GNNs) are effective for representing physical objects and have excelled in modeling 3D structures~\citep{schutt2018schnet, thomas2018tensor, DBLP:conf/nips/FuchsW0W20, satorras2021n, tholke2021equivariant}, as evidenced in applications like the open catalyst project~\citep{ocp_dataset, tran2022open}. To accommodate periodic materials, multi-graph edge construction~\citep{PhysRevLett.120.145301,yan2022periodic} and Fourier transforms to fractional coordinates~\citep{jiao2023crystal} were proposed to represent periodicity. In our work, we utilize Fourier transforms to achieve periodic transition invariance and constrain the additional lattice permutation invariance.


\textbf{Diffusion Generative Models.} 
Rooted in non-equilibrium thermodynamics theory~\citep{sohl2015deep}, diffusion models establish a link between data and prior distributions through forward and backward Markov chains~\citep{ho2020denoising}, achieving significant advancements in image generation~\citep{rombach2021highresolution, ramesh2022}. When integrated with equivariant GNNs, these models efficiently generate samples from invariant distributions, proving effective in tasks such as conformation generation~\citep{xu2021geodiff, shi2021learning}, ab initio molecule design~\citep{hoogeboom2022equivariant}, and protein generation~\citep{luo2022antigen}. DiffCSP distinguishes itself by simultaneously generating lattice and atom coordinates for crystals, utilizing a periodic-E(3)-equivariant denoising model~\cite{jiao2023crystal}. However, it has yet to fully realize E(3) equivariance based on periodic graph symmetry during its diffusion training process.

\section{Preliminaries}

\label{sec:notation}
\textbf{Crystal structures.} 
A 3D crystal structure is depicted as an endlessly repeating pattern of atoms in three-dimensional space, with the basic repeating entity known as a `unit cell'. This unit cell is defined by a triplet $\gM = (\mA, \mX, \mL)$, where $\mA = [\va_1, \va_2, \ldots, \va_n] \in \sR^{h \times n}$ symbolizes the one-hot encoded representations of atom types, $\mX = [\vx_1, \vx_2, \ldots, \vx_n] \in \sR^{3 \times n}$ comprises the atoms' Cartesian coordinates and $\mL = [\vl_1, \vl_2, \vl_3] \in \sR^{3 \times 3}$ represents the lattice matrix that indicates the repeating parameters of the unit cell. 
We represent periodic crystal structure as:
\begin{align}
    \{(\va_i',\vx_i')|\va_i'=\va_i,\vx_i'=\vx_i + \mL\vk, \forall\vk\in\sZ^{3\times 1}\},
\end{align}
where the $j$-th element of the integral vector $\vk$ denotes the integral 3D translation in units of $\vl_j$. 

\textbf{Fractional coordinate system.}
In crystallography, the fractional coordinate system is often used to represent the periodic nature of crystal structures~\cite{nouira2018crystalgan,kim2020generative,REN2021,hofmann2003crystal}. This system employs lattice vectors $(\vl_1, \vl_2, \vl_3)$ as coordinate bases, distinguishing it from the Cartesian system with its three orthogonal bases. A point in the fractional coordinate system, denoted by the vector $\vf=[f_1, f_2, f_3]^\top\in [0,1)^{3}$, corresponds to a Cartesian vector $\vx = \sum_{i=1}^3 f_i\vl_i$. All atomic coordinates in a cell compose $\mF\in[0,1)^{3\times n}$. This representation inherently maintains invariance to rotational and reflective transformations of the crystal structure. As described in \cite{mardia2000directional}, periodic data on each lattice base can be visualized as points on a circle, measured by angle value as depicted in Figure~\ref{fig:e3_vari} (e).

\textbf{Lattice parameters}
In crystallography, the lattice matrix $\mL$ can be converted to an invariant representations with three lattice lengths $\vl=[l_1, l_2, l_3]^\top$, where $l_i$ = $\|\vl_i\|_2$, and three lattice angles $\boldsymbol{\phi} = [\phi_{23}, \phi_{13}, \phi_{12}]$, where $\phi_{ij}$ is the angle between $\vl_i$ and $\vl_j$~\citep{hofmann2003crystal,luo2023towards}. This paper employs lattice parameters $\mC=[\vl, \boldsymbol{\phi}]\in \sR^{3 \times 2}$ instead of lattice matrix and represents the crystal by $\gM=(\mA,\mF,\mC)$. 


\textbf{Task definition.}
The CSP task entails predicting the lattice parameters $\mC$ and the fractional matrix $\mF$ for each unit cell, based on its chemical composition $\mA$. Specifically, this involves learning the conditional distribution $p(\mC, \mF \mid \mA)$.

\section{Methods}

This section initially outlines the symmetries inherent in crystal geometry, subsequently provides an overview of \model{}, and then introduces the joint equivariant diffusion process applied to $\mC$ and $\mF$, followed by the architecture of the denoising model.

\subsection{Symmetries of Crystal Structure Distribution}
\label{sec:symmetry}

The primary challenge of CSP lies in capturing the distribution symmetries of crystal structures. To tackle this, we define four key symmetries as representations within the distribution $p(\mC,\mF\mid \mA)$: composition permutation invariance, O(3) invariance, periodic translation invariance and lattice permutation invariance. Detailed definitions are provided as follows.


\begin{figure}[htbp]
  \centering
  \includegraphics[width=0.4\textwidth]{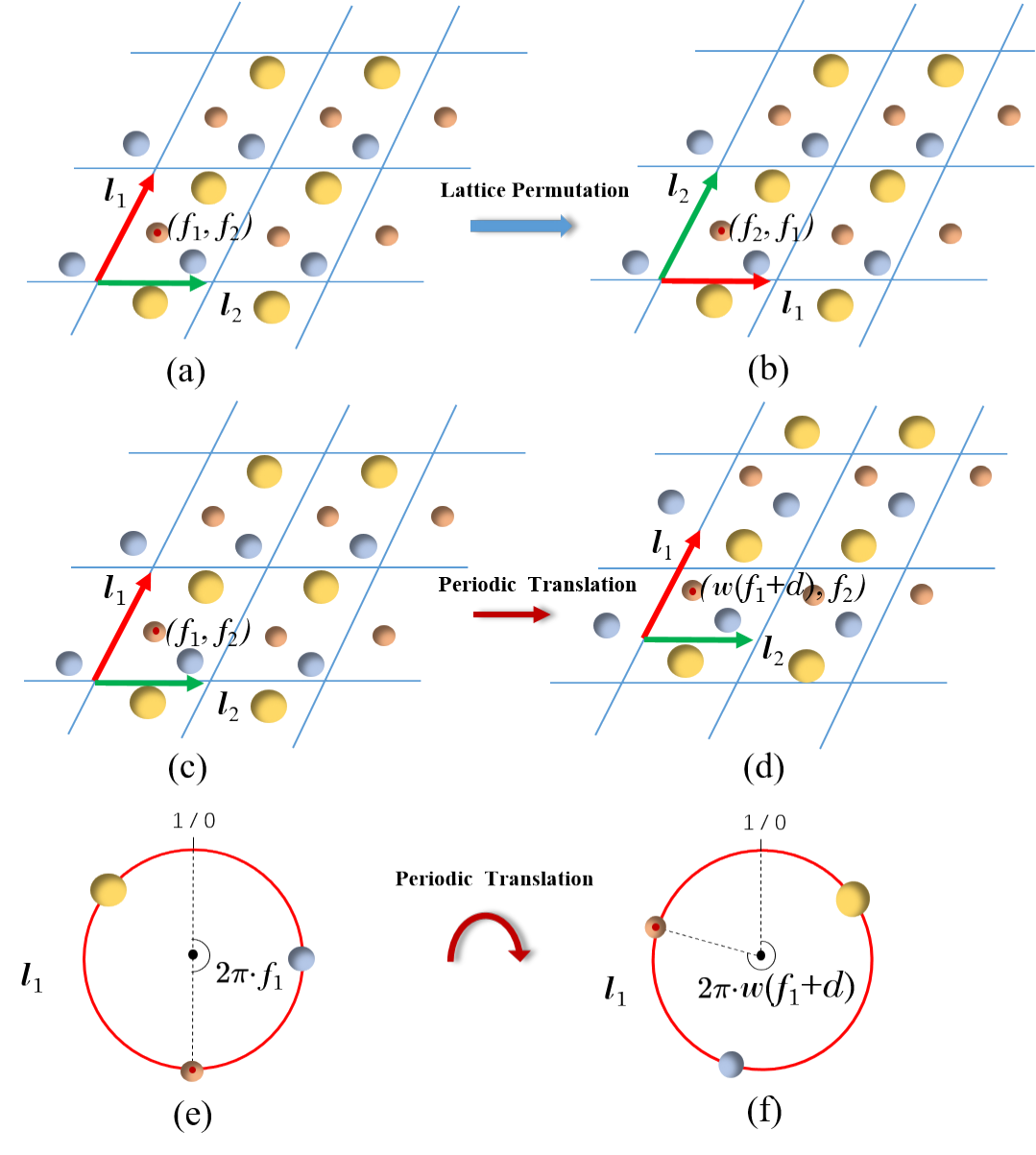}
  \caption{(a)$\rightarrow$(b): The lattice permutation of the lattice bases $\vl_1,\vl_2$. (c)$\rightarrow$(d): The periodic translation of the fractional coordinates $\vf_1, \vf_2$. (e)$\rightarrow$(f): The schematic diagram of the period translation represented as points on a circle. Both cases do not change the crystal structure. Here, the 2D crystal is used for better illustration.}
  \label{fig:e3_vari}
\end{figure}


\begin{definition}[{Composition Permutation Invariance}] 
\label{De:cpi}
For any permutation $\mP\in\mathrm{S}_n$, $p(\mC,\mF\mid \mA)=p(\mC,\mF\mP\mid \mA\mP)$, \emph{i.e.}, changing the order of atoms will not change the distribution, where S$_n$ represents the set of permutation matrices with dimensions $n\times n$.  
\end{definition}


\begin{definition} [O(3) Invariance]
\label{De:oi}
Given an transformation matrix $\mQ \in \sR^{3 \times 3}$ where $\mQ$ is any O(3) group element operated on $\mL$, the condition $p(\mC(\mQ\mL),\mF \mid \mA) = p(\mC(\mL),\mF \mid \mA)$ holds, indicating that the distribution remains invariant under any rotation or reflection applied to $\mL$, where $\mC(\cdot)$ is the function that translates a lattice matrix to lattice parameters.
\end{definition}

\begin{definition} [Lattice Permutation Invariance]
\label{De:lpi}
For any permutation $\mP\in\mathrm{S}_3$, $p(\mC,\mF\mid \mA)=p(\mP\mC,\mP\mF\mid \mA)$, \emph{i.e.}, changing the lattice base order will not change the distribution.   
\end{definition}

\begin{definition}[Periodic Translation Invariance]
\label{De:PTI}
For any translation $\rvt\in\R^{3\times1}$, $p(\mC, w(\mF + \rvt\vone^\top)\mid\mA)=p(\mC, \mF \mid\mA)$, where the function $w(\mF)=\mF - \lfloor\mF\rfloor \in [0,1)^{3\times n}$ returns the fractional part of each element in $\mF$, and $\vone\in\R^{3\times1}$ is a vector with all elements set to one. It explains that any periodic translation of $\mF$ will not change the distribution.  
\end{definition}


\begin{figure*}[htp]
  \centering
  \includegraphics[width=1.0\textwidth]{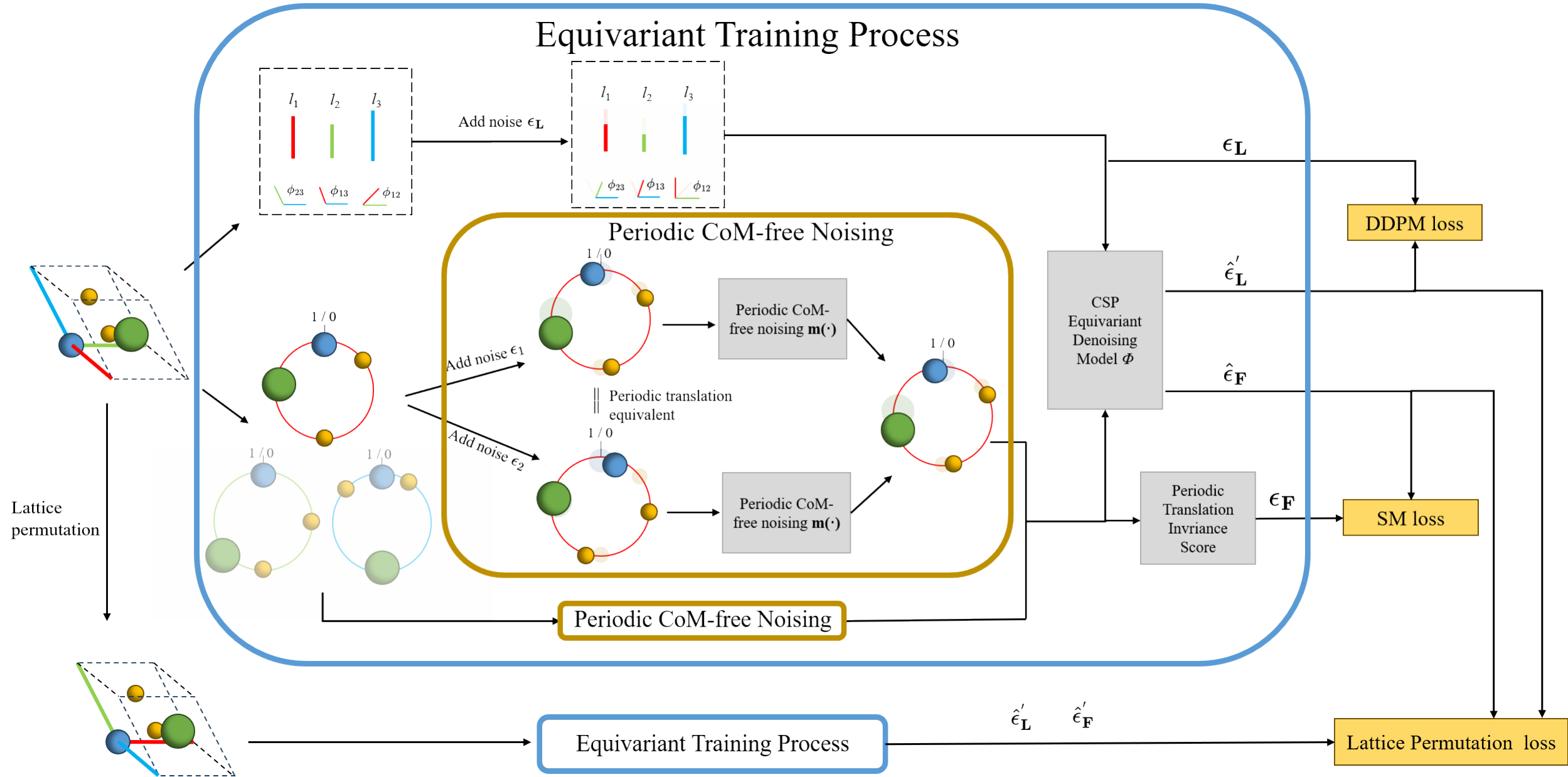}
  \vskip -0.1in
  \caption{Overview of training process in \model{}.}
  \label{fig:overview}
  \vskip -0.2in
\end{figure*}

Composition permutation invariance in generation is effectively achieved by GNNs as the foundational architecture~\citep{kipf2016variational}. 
According to previous work~\citep{jiao2023crystal}, employing the fractional system handles the O(3) invariance of crystals by ensuring O(3) invariance with respect to orthogonal transformations on the lattice matrix. Previous work~\citep{luo2023towards} further address the O(3) invariance of the lattice matrix by substituting it with lattice parameters $\mC$, as $\mC(\mQ\mL)=\mC(\mL)$ always holds for arbitrary $\mQ\in\text{O}(3)$. Consequently, our representation of crystals using both the fractional system and lattice parameters naturally satisfies O(3) invariance. Thus, we mainly focus on the lattice permutation and periodic translation invarance as shown in Figure~\ref{fig:e3_vari}.
For better demonstration, we utilize representation method in ~\cite{mardia2000directional} to show the periodic translation invariance. 
For details on the representation of lattice bases as circles in Figure~\ref{fig:e3_vari} (e) and (f), see Appendix~\ref{method:symme}.

\textbf{Comparing with other symmetry awareness generation method.} We notice that previous approaches~\citep{xie2021crystal, luo2023towards, jiao2023crystal} ignore the lattice permutation invariance, both for CSP task and for ab initio generation task. The ab initio generation method SymMat~\citep{luo2023towards} directly generates lattice parameters $\mC$ using a variational autoencoders from rand noise $\boldsymbol{\epsilon}$. However, it doesn't guarantee that the marginal distribution satisfies $p(\mC )=p(\mP\mC)$ for any $\mP\in\mathrm{S}_3$, which means that it doesn't guarantee the lattice permutation invariance. DiffCSP~\cite{jiao2023crystal} generates lattice matrix by diffusion model, however, as discussed in Section~\ref{sec:diffution_L}, its diffusion method lacks lattice permutation equivariance, impacting the final lattice distribution not invariant.
Our method is the first to realize this symmetry of crystal structure, and we will ablate the benefit in Section~\ref{sec:abla_stu}.


\subsection{An Overview of \model{}}
In our work, we implement \model{} by concurrently diffusing the $\mC$ and $\mF$ within the framework of DiffCSP~\cite{jiao2023crystal}. For a given atomic composition $\mA$, the intermediate states of $\mC$ and $\mF$ at any time step $t$ (where $0 \leq t \leq T$) are represented by $\gM_t$. \model{} orchestrates two distinct Markov processes: a forward diffusion that incrementally introduces noise into $\gM_0$, and a backward generation process that strategically samples from the prior distribution $\gM_T$ to reconstruct the initial data $\gM_0$. The implementation specifics are summarized in Algorithms~\ref{alg:train} and~\ref{alg:gen}.

In light of the symmetry discussed in Section~\ref{sec:symmetry}, the distribution restored from $\gM_T$ must meet invariance. This requirement is achieved if the prior distribution $p(\gM_T)$ exhibits invariance and the Markov transition $p(\gM_{t-1} | \gM_t)$ is equivariant, as established in previous literature~\cite{xu2021geodiff}. An equivariant transition implies $p(g\cdot \gM_{t-1} | g\cdot \gM_t)=p(\gM_{t-1} | \gM_t)$ for any transformation $g$ acting on $\gM$, as defined in Definitions~\ref{De:lpi}-\ref{De:PTI}.
Further explanations on how diffusion processes are applied to $\mC$ and $\mF$ are detailed subsequently.



\subsection{Diffusion on Lattice Parameters}
\label{sec:diffution_L}

Given that $\mC$ is a continuous variable with lattice lengths $\vl>0$ and lattice angles $\boldsymbol{\phi}\in (0, \pi)^{3}$, we exploit Denoising Diffusion Probabilistic Model (DDPM)~\citep{ho2020denoising} with prepossessing of $\mC$ to accomplish the generation.
As detailed in Appendix~\ref{ddpm_L}, such preprocessing projects the definition domain of  $\mC$ onto $\sR^{3\times 2}$, and hereinafter the notation $\mC$ refers to the projected lattice parameters.

\subsubsection{Generation}
\label{sec:L_gen}
We define the generation process that progressively diffuses the Normal prior $p({\mC}_T)$ towards stable crystal lattice distribution $p({\mC}_0)$ by:
\begin{align}
\label{eq:ddpm1}
    p({\mC}_{t-1} | \gM_t) &= \gN({\mC}_{t-1} | \mu(\gM_t),\sigma^2 (\gM_t)\mI), 
\end{align}
where $\mu(\gM_t) = \frac{1}{\sqrt{\alpha_t}}\Big(\mC_t - \frac{\beta_t}{\sqrt{1-\bar{\alpha}_t}}\hat{\vepsilon}_\mL(\gM_t,t)\Big),\sigma^2(\gM_t) $$= \beta_t \frac{1-\bar{\alpha}_{t-1}}{1-\bar{\alpha}_t}$. The denoising term $\hat{\vepsilon}_\mL(\gM_t,t)\in\R^{3\times2}$ is predicted by the neural network model $\phi(\mC_t,\mF_t,\mA,t)$ detailed in Section~\ref{sec:model}.

As the prior distribution $p(\mC_T)=\gN(0,\mI)$ is already lattice permutation invariant, we require the generation process in Eq.~(\ref{eq:ddpm1}) to be lattice permutation equivariant, which is formally stated below, and give a proof in Appendix~\ref{proof:pro_45}. 
\begin{proposition}
\label{prop:lpi}
The marginal distribution $p(\mC_0)$ by Eq.~(\ref{eq:ddpm1}) is lattice permutation invariant if $\hat{\vepsilon}_\mL(\gM_t,t)$ is lattice permutation equivariant, namely $ \hat{\vepsilon}_\mL(\mP\mC_t,\mP\mF_t,\mA,t) =\mP\hat{\vepsilon}_\mL(\mC_t,\mF_t,\mA,t), \forall\mP\in\mathrm{S}_3$.
\end{proposition}

\subsubsection{Training}
\label{sec:L_train}
We define the forward process as one that gradually diffuses $\mC_0$ towards a Normal prior, represented by $p(\mC_T) = \gN(0, \mI)$. This process is defined through the conditional probability $q(\mC_t | \mC_{t-1})$, which is formulated based on the initial distribution:
\begin{align}
\label{eq:ini_dis}
q(\mC_t | \mC_{0}) &= \gN\Big(\mC_t | \sqrt{\bar{\alpha}_t}\mC_{0}, (1 - \bar{\alpha}_t)\mI\Big),  
\end{align}
where $\beta_t\in(0,1)$ controls the variance, and $\bar{\alpha}_t = \prod_{s=1}^t \alpha_t= \prod_{s=1}^t (1-\beta_t)$ is valued in accordance to the cosine scheduler~\citep{nichol2021improved}.

To train the denoising model $\phi$, we initiate by sampling $\vepsilon_\mL \sim \gN(0, \mI)$ and reparameterize $\mC_t=\sqrt{\bar{\alpha}_t}\mC_{0} + \sqrt{1 - \bar{\alpha}_t}\vepsilon_\mL$ based on Eq.~(\ref{eq:ini_dis}). The training goal is then established by minimizing the $\ell_2$ loss between $\vepsilon_\mL$ and its estimate $\hat{\vepsilon}_\mL$:
\begin{align}
\gL_\mC&=\E_{\vepsilon_\mL\sim\gN(0,\mI), t\sim\gU(1,T)}[\|\vepsilon_\mL - \hat{\vepsilon}_\mL(\gM_t,t)\|_2^2].
\end{align}

To satisfy proposition~\ref{prop:lpi}, we introduce an additional loss value as a penalty term during the training process, detailed as follows:
\begin{align}
\gL_{p\mC} &=
\E[\|\hat{\vepsilon}_\mL(\mP\mC_t, \mP\mF_t, \mA, t) - \mP\hat{\vepsilon}_\mL(\mC_t, \mF_t, \mA, t)\|_2^2],
\end{align}

 where the expectation is taken with respect $t\sim\gU(1,T)$ and $\mP \sim \mathcal{U} (\mathrm{S}_3)$. After sufficient training, the generation process will satisfy lattice permutation invariance as stated in Proposition~\ref{prop:lpi}. Our ablation experiments demonstrate the significant performance of this method, and the learning curve in Appendix~\ref{learn_curves} shows that the difficulty of learning is greatly reduced compared with DiffCSP~\citep{jiao2023crystal}.
 
 \textbf{Comparing with the Method of Encoding the Equivariance to Denoising Model}. We discover that the Frame Average (FA) method~\citep{puny2021frame}, employing a unified, hard-constraint approach for equivariant neural networks, also satisfies Proposition~\ref{prop:lpi}. We provide implementation details in Appendix~\ref{appen:fa}. However, our experimental findings in Section~\ref{sec:abla_stu} reveal that the computational burden of finite group operations required by FA renders it impractical for iterative models like diffusion models. In contrast, our method significantly enhances computational efficiency and accuracy by simply incorporating additional loss values during training. 

\subsection{Diffusion on Fractional Coordinates}
\label{sec:frac}
Combining Score-Matching (SM) based framework with Wrapped Normal (WN) distribution (SMWN), as proposed in \cite{jiao2023crystal}, for generating fractional coordinates proves advantageous due to the periodicity and [0,1) constraint of these coordinates. Based on SMWN method, we propose an innovative noising algorithm to meet periodic translation equivariance, detailed as follows:


\subsubsection{Generation}
\label{generation for frac}

In the generation process, we first initialize $\mF_T$ from the uniform distribution $\gU(0,1)$, which is periodic translation invariant. With the denoising term $\hat{\vepsilon}_\mF(\gM_t, t)$ predicted by $\phi(\mC_t,\mF_t,\mA,t)$ to model the data score $\nabla_{\mF_t}\log p(\mF_t)$, we combine the ancestral predictor with the Langevin corrector used in DiffCSP~\cite{jiao2023crystal} to sample $\mF_{0}$. Specifically, this method can be simply viewed as progressively sampling from the wrapped normal distribution \(p(\mF_{t-1}|\gM_t)\) at each time step $t$, where the mean of the wrapped normal is a function of \(\hat{\vepsilon}_\mF\), with the detailed formula provided in Eq.~(\ref{eq:F_generation}) of the Appendix~\ref{proof:pro_46}. To ensure that \(p(\mF_{t-1}|\gM_t)\) satisfies periodic translation equivariance, in accordance with \cite{jiao2023crystal}, \(\hat{\vepsilon}_\mF\) must meet the periodic translation invariance:
\begin{align}
\label{eq:peri_trans}
\hat{\vepsilon}_\mF(\mC_t, \mF_t, \mA, t)=\hat{\vepsilon}_\mF(\mC_t, w(\mF_t+\rvt\vone^\top), \mA, t), 
\end{align}
where $\forall\rvt\in\R^3$ and the truncation function $w(\cdot)$ is already defined in Definition~\ref{De:PTI}. We will ensure that the model output conforms to this property in Section~\ref{sec:model} to guarantee the equivariance of generation. 

Similarly, the data score $\nabla_{\mF_t}\log p(\mF_t)$ must adhere to periodic translation invariance. The accurate estimation of this score, set as the training target for $\hat{\vepsilon}_\mF(\gM_t, t)$, represents the primary challenge we will tackle in Section~\ref{training for frac}.

In addition, we require the generation process to be lattice permutation equivariant, which is formally stated below, provided a proof in Appendix~\ref{proof:pro_46}:
\begin{proposition}
\label{prop:lpif}
The marginal distribution $p(\mF_0)$ is lattice permutation invariant if $\hat{\vepsilon}_\mF(\gM_t,t)$ is lattice permutation equivariant, namely $ \hat{\vepsilon}_\mF(\mP\mC_t,\mP\mF_t,\mA,t)=\mP \hat{\vepsilon}_\mF(\mC_t,\mF_t,\mA,t), \forall\mP\in\mathrm{S}_3$. 
\end{proposition}

\subsubsection{Training}
\label{training for frac}

During the forward process, SMWN samples each column of $\vepsilon\in\R^{3\times n}$ from wrapped normal distribution $\gN_w(0,\sigma_t\mI)$, and then acquire $\mF_t = w(\mF_0 + \vepsilon)$, where $\gN_w(0,\sigma_t^2\mI)$ denotes the probability density function(PDF) of WN distribution with mean 0, variance $\sigma_t^2$ and period 1, $\sigma_t$ is the noise magnitude level and $\sigma_1 < \sigma_2 < \ldots \sigma_T$. According to the feature of WN, if $\sigma_T$ is sufficiently large, $p(\mF_T)$ approaches a uniform distribution $\gU(0,1)$ which is desirable for generation. Our training target is:
\begin{align}
\label{eq:training_target}
\hat{\vepsilon}_\mF(\gM_t,t) \rightarrow \nabla_{\mF_t}\log q(\mF_t).
\end{align}
The pivotal challenge is how to accurately obtain the score matrix $\nabla_{\mF_t}\log q(\mF_t)$ to maintain the periodic translation invariance, a feature not guaranteed by the conventional diffusion framework. For instance, DiffCSP~\cite{jiao2023crystal} employs the ordinary denoising score matching~\cite{vincent2011connection} training objective to estimate the score: 
\begin{align}
\label{eq:ordinary_target}
\begin{aligned}
\gL_\mF &= \E_{\mF_0\sim q(\mF_0),\mF_t\sim q(\mF_t | \mF_0), t\sim\gU(1,T)}\\
&\big[\lambda_t\|\nabla_{\mF_t}\log q(\mF_t | \mF_0)-\Tilde{\mS}\|_2^2\big],
\end{aligned}
\end{align}
where $\Tilde{\mS}$ is the estimate of $\nabla_{\mF_t}\log q(\mF_t)$ and $\lambda_t$ is the weight of loss. The core issue arises from the dataset distribution $q(\mF_0)$, which typically does not exhibit the same periodic translation invariance as the ground truth distribution $p(\mF_0)$ because the dataset usually does not contain all the samples $w(\mF_0+\rvt\vone^\top)$. Consequently, $\Tilde{\mS}$ cannot be guaranteed to be periodic translation invariant. For illustration, consider a dataset with only one sample $\Tilde{\mF_0}$, the estimate of score will be:
\begin{align}
\label{eq:example}
\begin{aligned}
\Tilde{\mS} &= \nabla_{\mF_t}\log q(\mF_t | \mF_0=\Tilde{\mF_0}) \\
&=\nabla_{\mF_t}\log \gN_w(\mF_t|\Tilde{\mF_0},\sigma_t^2\mI)\\
 &=\nabla_\vepsilon\log \gN_w(\vepsilon|0,\sigma_t^2\mI),
\end{aligned}
\end{align}
and obviously
\begin{align}
\label{eq:score_wrong}
\begin{aligned}
&\quad\nabla_\vepsilon\log \gN_w(\vepsilon|0,\sigma_t^2\mI)\\
&\neq \nabla_{w(\vepsilon+\rvt\vone^\top)} \log \gN_w\big(w(\vepsilon+\rvt\vone^\top)|0,\sigma_t^2\mI\big),
\end{aligned}
\end{align}
which means the predicted score not invariant. Numerous studies~\citep{luo2023towards,luo2021predicting,niu2020permutation,jin2023dsmbind} have also indicated that for ensuring equivariance, the score matrix should be determined more cautiously.

A potential solution to this issue is to augment the dataset using periodic translation operations to better align $q(\mF_0)$ with the invariant distribution $p(\mF_0)$. However, this approach demands a significant amount of training time due to the periodic translation group being a Lie group with infinitely many elements. 

We propose `Periodic CoM-free Noising', a new noising method that ensures the noise added to $q(\mF_0)$ results in periodic translation invariant score as closely aligned as possible to the score achieved by the original noise added to $p(\mF_0)$. The method is based on the following statements:
$\nabla_{\mF_t}\log q(\mF_t)$ is periodic translation invariant if  $\nabla_{\mF_t}\log q(\mF_t | \mF_0)$ is periodic translation invariant: 
\begin{align}
\label{eq:scorepti}
\begin{aligned}
 &\quad\nabla_{\mF_t}\log q(\mF_t | \mF_0) \\
 &= \nabla_{w(\mF_t+\rvt\vone^\top)}\log q(w(\mF_t+\rvt\vone^\top)|\mF_0),
\end{aligned}
\end{align} 
where $\forall\rvt\in\R^3$. 

The noising method is equivalent to operating on $\nabla_{\mF_t}\log q(\mF_t | \mF_0)$. Therefore, we first focus on achieving $\nabla_{\mF_t}\log q(\mF_t | \mF_0)$ that meets the periodic translation invariance, followed by adjusting the score numerically for more accurate training results.

\textbf{Periodic CoM-free Noising.} In order to satisfy Eq.(\ref{eq:scorepti}), we adopt a parameterization scheme for $\nabla_{\mF_t}\log q(\mF_t | \mF_0)$ as follows:
\begin{align}
\label{eq:PCN}
\begin{aligned}
\nabla_{\mF_t}\log q(\mF_t | \mF_0) &= \nabla_{\bar{\mF}}\log \mathcal{N}_w(\bar{\mF} | \mF_0, \sigma_t^2 \mathbf{I})\\
&=\nabla_{\bar{\vepsilon}} \log \mathcal{N}_w(\bar{\vepsilon} | 0, \sigma_t^2 \mathbf{I}),
\end{aligned}
\end{align}
where 
\begin{align}
\bar{\mF} &= w(\mF_0 + \bar{\vepsilon}),\\
\label{eq:m_invariant}
\bar{\vepsilon} &= \vm(\vepsilon) = \vm(w(\vepsilon + \rvt\vone^\top)),\forall\rvt\in\R^3.
\end{align}
Here, we introduce a noise conversion function $\vm(\cdot)$ to map all the fractional coordinate matrices that are periodic translation equivalent with $\mF_t$ to a unique matrix $\bar{\mF}$. This addresses the requirement of periodic translation invariance of score. Consequently, we can employ the ordinary score calculation method, specifically the score of anisotropic WN here, to compute $\nabla_{\bar{\mF}}\log q(\bar{\mF} | \mF_0)$ as a substitute for the required score.

The key of Periodic CoM-free Noising is to design the specific function $\vm:w(\vepsilon+\rvt\vone^\top) \rightarrow \bar{\vepsilon}$. We note that the CoM-free systems in molecular conformation generation~\cite{xu2022geodiff} solve similar problem in translation invariance. However, the Center of Mass(CoM) of periodic data cannot be simple computed as mean value of data~\cite{bai2008calculating}. Similar to the idea of CoM-free systems, we utilize the concept of ``mean angle" from \citep{mardia2000directional} to construct $\vm(\cdot)$ as a periodic CoM-free function. Specifically, we denote $\vepsilon=[\vepsilon_1, \vepsilon_2, \ldots, \vepsilon_n]$ and fomulate:
\begin{align}
\label{eq:pcom}
\left\{
\begin{aligned}
\vm(\vepsilon) &= w\big(\vepsilon-\frac{\atan2 \left(\bar{\vy}(\vepsilon), \bar{\vx}(\vepsilon)\right)}{2\pi}\vone^\top\big),\\
\bar{\vy}(\vepsilon) &=\frac1n\sum_{i=0}^n\sin{(2\pi\vepsilon_{i})},\\
\bar{\vx}(\vepsilon) &=\frac1n\sum_{j=0}^n\cos{(2\pi \vepsilon_{i})}.
\end{aligned}
\right.
\end{align}
Intuitively, as shown in Figure 3, the function transform periodic data of each lattice axis to angle data on a circle, and then subtract all the data by the periodic CoM. Consequently, it maps all equivalent periodic data to the same representation, and addresses the periodic translation invariance. We provide a proof in Appendix~\ref{proof:p_com}.

\begin{figure}[htbp]
  \centering
  \includegraphics[width=0.3\textwidth]{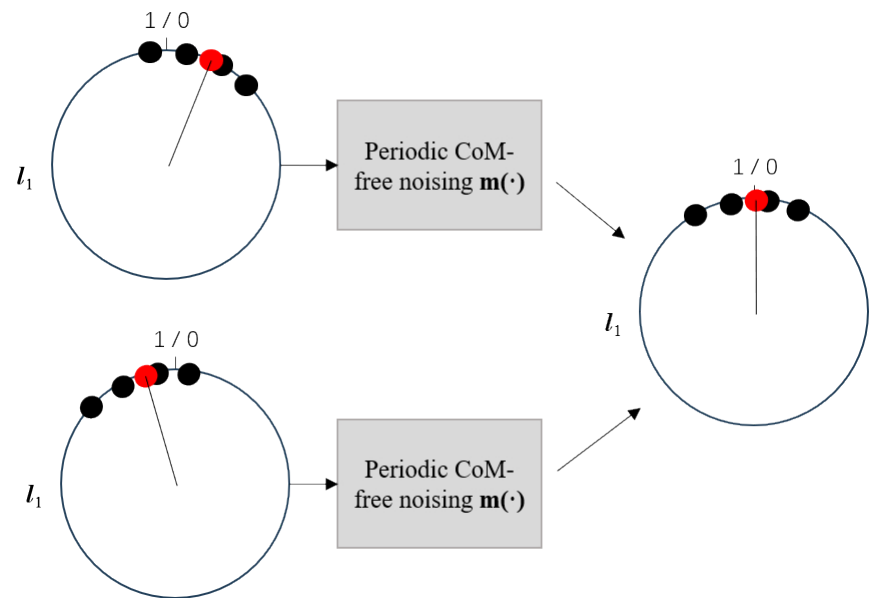}
  \caption{The illustration of periodic translation invariance with periodic CoM-free function.}
  \label{fig:com_free_func}
  \vskip -0.1in
\end{figure}

After implementing the algorithm, we substituted the $\nabla_{\mF_t}\log q(\mF_t | \mF_0)$ in the denoising score matching training objective, as defined in Eq.(\ref{eq:ordinary_target}), with Eq.(\ref{eq:PCN}). This change led to significant performance improvements in our ablation study and excellent training convergence demonstrated in Appendix~\ref{learn_curves}. However, with $\bar{\mF} = [\bar{\vf_1}, \bar{\vf_2}, \ldots, \bar{\vf_n}]$ and $\mF_t = [\vf_1, \vf_2, \ldots, \vf_n]$, we identified two points that still require enhancement: \textbf{1.} The marginal distribution\footnote{We focus on evaluating the score of marginal distribution instead of the joint distribution to better align with the origin noise addition techniques employed in SDEs.} $q(\bar{\vf_i} | \mF_0)$ does not simply satisfy $\gN_w(0, \sigma_t^2\mI)$, so it is necessary to re-evaluate its distribution to recalculate $\nabla_{\bar{\mF}}\log q(\bar{\mF} | \mF_0)$. \textbf{2.} The generation process is designed for non periodic CoM-free system, while our method now simply use periodic CoM-free score $\nabla_{\bar{\mF}}\log q(\bar{\mF})$ to replace corresponding $\nabla_{\mF_t}\log q(\mF_t)$. A more rigorous implementation of probabilistic modeling is warranted to establish a refined connection between the score. We provide solutions below.

\textbf{Von Mises Simulation.} We denote $\bar{\vepsilon}=[\bar{\vepsilon}_1, \bar{\vepsilon}_2, \ldots, \bar{\vepsilon}_n]$. To evaluate the probability density function (PDF) $q(\bar{\vf_i}|\mF_0)$, we recognize that since $\bar{\mF} = w(\mF_0 + \bar{\vepsilon})$, the task can be reframed as estimating the marginal PDF of $\bar{\vepsilon} = \vm(\vepsilon)$, specifically $p(\bar{\vepsilon}_i)$, where $\vepsilon \sim \gN_w(0, \sigma_t^2\mI)$.  However, directly obtaining the formula of $p(\bar{\vepsilon}_i)$ is challenging due to the complexity of $\vm(\cdot)$ and WN. For simplicity, we utilize the Von Mises distribution~\citep{gatto2007generalized} which is commonly used in dealing with circular distribution problems to simulate the $p(\bar{\vepsilon}_i)$, and use the Monte Carlo method to obtain its parameters. More details is list in Appendix~\ref{sec:von_mises}. Consequently, $\nabla_{\bar{\mF}}\log q(\bar{\mF} | \mF_0)=[\bar{\vc}_1,\bar{\vc}_2 \ldots \bar{\vc}_n]$ can be expressed using the formula of score of Von Mises:
\begin{align}
\label{eq:vonmises}
\begin{aligned}
\bar{\vc}_i &= \nabla_{\bar{\vf}_i}\log q(\bar{\vf}_i | \mF_0) = \nabla_{\bar{\vepsilon}_i}\log p(\bar{\vepsilon}_i) \\
&=-2\pi\cdot\kappa(n,\sigma_t)\cdot \sin(2\pi\bar{\vepsilon}_i),
\end{aligned}
\end{align}
where $\kappa(n,\sigma_t)$ is the parameter of Von Mises distribution obtained by Monte Carlo method.  Consequently, using the revised $\nabla_{\bar{\mF}}\log q(\bar{\mF} | \mF_0)$ allows for a more accurate estimation of the invariant score $\nabla_{\bar{\mF}}\log q(\bar{\mF}) = [\bar{\vs}_1, \bar{\vs}_2, \ldots \bar{\vs}_n]$. 

\textbf{Probabilistic Modeling Process.} For the second point, we propose a novel probabilistic modeling process inspired by ~\citep{luo2023towards}. Denoting the score of non periodic CoM-free system as $\nabla_{\mF_t}\log q(\mF_t)=[\vs_1,\vs_2 \ldots \vs_n]$, we consider $\vs_i$ as a function of all the corresponding CoM-free data namely $\{\bar{\vf}_1, \bar{\vf_2}, \ldots, \bar{\vf}_n\}$ aiming to address periodic translation invariance. From the chain rule of derivatives, we can approximate the score $\nabla_{\mF_t}\log q(\mF_t)$ by the score of 
 CoM-free system namely $\nabla_{\bar{\mF}}\log q(\bar{\mF})$:
\begin{align}
\label{eq:scorebb}
\begin{aligned}
 \vs_j &=\sum_{i=0}^n \nabla_{\bar{\vf}_i}\log q(\bar{\vf}_i) \cdot \nabla_{\vf_j} \bar{\vf_i},\\
 &=\sum_{i=0}^n\bar{\vs}_i \cdot \nabla_{\vf_j} \bar{\vf_i}.
\end{aligned}
\end{align}
We previously established that $\bar{\vs}_i$ is invariant, but $\nabla_{\vf_j} \bar{\vf_i}$ might destroy the periodic translation invariance. Fortunately, we can first transform $\nabla_{\vf_j} \bar{\vf_i}$ to the $j$-colomn of $\nabla_\vepsilon \bar{\vepsilon}_i$ by their definition, and then strictly prove the following statements in Appendix~\ref{apd:mgrad}:
\begin{proposition}
\label{prop:ptiexp}
 $\nabla_\vepsilon{\bar{\vepsilon}_i}=\nabla_\vepsilon\big( \vm({\vepsilon})[:,i]\big)$ is periodic translation invariance, where $\vm({\vepsilon})[:,i]$ is the $i$-th column of $\vm({\vepsilon})$. In other words, $\nabla_\vepsilon\big( \vm({\vepsilon})[:,i]\big)=\nabla_{w(\vepsilon+\rvt\vone^\top)}\big( \vm({w(\vepsilon+\rvt\vone^\top)})[:,i]\big)$.
 Thus $\nabla_{\mF_t}\log q(\mF_t)$ by Eq.(\ref{eq:scorebb}) is periodic translation invariance. 
\end{proposition}
While $\vm(\bar{\vepsilon})=\vm(\vepsilon)$ holds, we can reformulate Eq.(\ref{eq:scorebb}) using Proposition~\ref{prop:ptiexp} as:
\begin{align}
\label{eq:score_after}
  \vs_j &=\sum_{i=0}^n\bar{\vs}_i \cdot \nabla_{\bar{\vepsilon}_j} \big(\vm({\bar{\vepsilon}})[:,i]\big), 
\end{align}
where $\nabla_{\bar{\vepsilon}_j} \big(\vm({\bar{\vepsilon}})[:,i]\big)$ is the $j$-column of $\nabla_{\bar{\vepsilon}} \big( \vm({\bar{\vepsilon}})[:,i]\big)$. This indicates that to compute the adjusted score $\nabla_{\mF_t}\log q(\mF_t)$, an additional invariant parameter $\bar{\vepsilon}$ is required. This parameter can be predicted by the model $\phi$. 

\textbf{Put things together.} Finally, we consolidate our approach to construct the training objective to approximate $\nabla_{\bar{\mF}}\log q(\bar{\mF})$ and the expected $\bar{\vepsilon}$:
\begin{align}
\begin{aligned}
\gL_s &=\E_{\mF_0 \sim q(\mF_0),\vepsilon \sim \gN_w(0, \sigma_t \mI),t\sim\gU(1,T)}\\
&[\| \nabla_{\bar{\mF}}\log q(\bar{\mF} | \mF_0)  - s_\theta(\gM_t,t)\|_2^2],
\end{aligned}
\end{align}
\begin{align}
\begin{aligned}
\gL_F &=\E_{\mF_0 \sim q(\mF_0),\vepsilon \sim \gN_w(0, \sigma_t \mI),t\sim\gU(1,T)}\\
&[\|{\bar{\vepsilon}} - F_\theta(\gM_t,t)\|_2^2],
\end{aligned}
\end{align}
where $s_\theta(\gM_t,t)$ and $F_\theta(\gM_t,t)$ are directly predicted by the model $\phi$, and $\nabla_{\bar{\mF}}\log q(\bar{\mF} | \mF_0)$ is calculated by Eq.(\ref{eq:vonmises}), meanwhile ${\bar{\vepsilon}}=\vm(\vepsilon)$. And we can finally derive $\hat{\vepsilon}_\mF$ in Section~\ref{generation for frac} by replacing $\nabla_{\bar{\mF}}\log q(\bar{\mF})$ with $s_\theta(\gM_t,t)$ and $\bar{\vepsilon}$ with $F_\theta(\gM_t,t)$ in Eq.(\ref{eq:score_after}).

In addition, to satisfy Proposition \ref{prop:lpif}, we design the permutation loss similar to the method in Section \ref{sec:diffution_L}:
\begin{align}
\gL_{ps} &=
\E[\|s_\theta(\mP\mC_t, \mP\mF_t, \mA, t) - \mP s_\theta(\mC_t, \mF_t, \mA, t)\|_2^2],\\
\gL_{pF} &=
\E[\|F_\theta(\mP\mC_t, \mP\mF_t, \mA, t) - \mP F_\theta(\mC_t, \mF_t, \mA, t)\|_2^2],
\end{align}
where the expectation is taken with respect $t\sim\gU(1,T)$ and $\mP \sim \mathcal{U} (\mathrm{S}_3)$.

\begin{table*}[!tbp]
\vskip -0.1in
  \centering
  \caption{Results on stable structure prediction task. The results of baseline methods are from Jiao~\cite{jiao2023crystal}}
  \resizebox{0.9\linewidth}{!}{
  \small
   \setlength{\tabcolsep}{3.2mm}
    \begin{tabular}{p{1.2cm}ccccccccc}
    \toprule
 & \multicolumn{2}{c}{Perov-5 } & \multicolumn{2}{c}{MP-20} & \multicolumn{2}{c}{MPTS-52} \\
            & Match rate$\uparrow$ & RMSE$\downarrow$  & Match rate$\uparrow$ & RMSE$\downarrow$ & Match rate$\uparrow$ & RMSE$\downarrow$ \\
    \midrule
    \midrule
        {RS} 
           & 36.56  & 0.0886  & 11.49  & 0.2822 & 2.68 &	0.3444  \\
    \midrule
        {BO}  & 55.09  & 0.2037  & 12.68  & 0.2816 & 6.69 &	0.3444 \\
    \midrule

        {PSO}  & 21.88  & 0.0844  & 4.35  & 0.1670 & 1.09 &	0.2390  \\
    \midrule
    \midrule
    \multirow{1}[1]{*}{P-cG-SchNet}    & 48.22  & 0.4179   & 15.39 & 0.3762 & 3.67 & 0.4115   \\
    \midrule
    {CDVAE}  & 45.31  & 0.1138  & 33.90  & 0.1045 & 5.34 & 0.2106 \\
    \midrule
    {DiffCSP}  & 52.02  & 0.0760  & 51.49 &	0.0631 & 12.19 & 0.1786 \\
    \midrule
    \midrule
    {\model{}}   & \textbf{52.02}  & \textbf{0.0707}  & \textbf{57.59} &\textbf{0.0510} & \textbf{14.85} & \textbf{0.1169} \\
        
    \bottomrule
    \end{tabular}%
    }
\vskip -0.1in
  \label{tab:oto}%
\end{table*}%


\subsection{The Architecture of the Denoising Model}
\label{sec:model}

In this subsection, we outline the specific design of the denoising model $\phi(\gM_t)$, focusing on how it computes the three denoising terms: $\hat{\vepsilon}_\mL, s_\theta, F_\theta$. For simplicity, the subscript $t$ is omitted in this discussion.

The model begins by integrating the atom embeddings $f_{\mathrm{atom}}(\boldsymbol{A})$ with sinusoidal time embeddings $f_{\mathrm{time}}(t)$ to generate the initial node features $\mH=\varphi_{\mathrm{in}}(f_{\mathrm{atom}}(\boldsymbol{A}), f_{\mathrm{time}}(t))$. We then describe the message passing mechanism from node $j$ to node $i$ in the l-th layer of the network.
\begin{align}
\vm_{ij}^{(l)} &=\varphi_m(\boldsymbol{h}_i^{(l-1)},\boldsymbol{h}_j^{(l-1)},\mC,\psi_{\mathrm{FT}}(\boldsymbol{f}_j-\boldsymbol{f}_i)) \\
\vh_{i}^{(l)} &=\boldsymbol{h}_i^{(l-1)}+\varphi_h(\boldsymbol{h}_i^{(l-1)},\sum_{j=1}^n\boldsymbol{m}_{ij}^{(l)}), 
\end{align}
 where $\varphi_m$ and $\varphi_h$ are MLPs, and The function $\psi_{\text{FT}}:(-1,1)^{3}\rightarrow [-1,1]^{3\times K}$ is Fourier Transformation of the relative fractional coordinate $\vf_j-\vf_i$ to address periodic translation invariance according to \citep{jiao2023crystal}. 
 
 After $S$ layers of message passing, we get the graph-level denoising term as:
 \begin{align}
\label{eq:ln}
    \hat{\vepsilon}_\mL &= \varphi_\mL\Big(\frac{1}{n}\sum_{i=1}^n \vh_i^{(S)}\Big), 
\end{align}
 
and the node-level denoising terms as
\begin{align}
\label{eq:fs}
    s_\theta[:,i],F_\theta[:,i] &= \varphi_s(\vh_i^{(S)}),\varphi_F(\vh_i^{(S)})
\end{align}
 
where $\varphi_{\mL},\varphi_{s},\varphi_{F}$ are MLPs.

\section{Experiments}
In this section, we evaluate the performance of \model{} on diverse tasks, by showing the capability of generating high-quality structures of different crystals in Section~\ref{sec:stable}. Ablations in Section~\ref{sec:abla_stu} show the necessity of each designed component. We further exhibit the capability of \model{} in the ab initio generation task in Appendix~\ref{sec:ab_initio_gen}.
\subsection{Stable Structure Prediction Results}
\label{sec:stable}
\textbf{Datasets.} 
Experiments are carried out on three datasets, each varying in complexity. The \textbf{Perov-5} dataset~\citep{castelli2012new, castelli2012computational} comprises 18,928 perovskite materials, characterized by their analogous structural configurations. Notably, each structure within this dataset features a unit cell containing 5 atoms.
The dataset \textbf{MP-20} comprises 45,231 stable inorganic materials curated from the Material Projects\cite{jain2013commentary}. This dataset predominantly includes materials that are experimentally generated and contain no more than 20 atoms per unit cell. In addition, \textbf{MPTS-52} represents a more challenging extension of MP-20, encompassing 40,476 structures with up to 52 atoms per cell. These structures are organized based on the earliest year of publication in the literature. For datasets such as Perov-5, and MP-20, we adhere to a 60-20-20 split for training, validation, and testing, respectively, aligning with the methodology of~\citet{jiao2023crystal}. Conversely, for MPTS-52, we allocate 27,380 entries for training, 5,000 for validation, and 8,096 for testing, arranged in chronological order.

\textbf{Baselines.} This study contrasts two categories of preceding research. The initial category adopts a predict-optimize approach, initially training a property predictor, followed by employing optimization algorithms for identifying optimal structures. Following~\citet{cheng2022crystal}, we use MEGNet~\citep{chen2019graph} for formation energy prediction. For optimization, we select Random Search (\textbf{RS}), Bayesian Optimization (\textbf{BO}), and Particle Swarm Optimization (\textbf{PSO}), each conducted over 5,000 iterations. The second category revolves around deep generative models. In line with modifications by~\citet{xie2021crystal}, we employ cG-SchNet~\citep{gebauer2022inverse}, integrating SchNet~\citep{schutt2018schnet} as its core and incorporating ground-truth lattice initialization to encode periodicity, resulting in the \textbf{P-cG-SchNet} model. 
Another baseline, \textbf{CDVAE}~\citep{xie2021crystal}, which is a VAE-based approach for crystal generation, predicts lattice and initial composition, and then optimizes atom types and coordinates using annealed Langevin dynamics. Following the method by~\citep{jiao2023crystal},we adapt CDVAE for the CSP task.
\textbf{DiffCSP}~\citep{jiao2023crystal}, a diffusion method, learns stable structure distributions, incorporating translation, rotation, and periodicity, effectively modeling material systems.

\textbf{Evaluation metrics.} Adhering to established protocols~\citep{xie2021crystal}, we assess performance by comparing predicted candidates against ground-truth structures. For each test set structure, we generate one samples with identical composition, considering a match if any sample aligns with the ground truth under pymatgen's StructureMatcher class metrics~\citep{ong2013python}, with ltol=0.3, angle\_tol=10, stol=0.5. The \textbf{Match rate} reflects the ratio of matched structures in the test set. \textbf{RMSE} is computed between the ground truth and the closest matching candidate, normalized by $\sqrt[3]{V/n}$ where $V$ represents the lattice volume, and averaged across matched structures. 


\textbf{Results.} 
Table~\ref{tab:oto} presents the following key insights:
\textbf{1.} Optimization approaches exhibit low Match rates, indicating the challenging nature of pinpointing optimal structures within the expansive search space.
\textbf{2.} Our method outperforms other generative approaches, which underscore our method's effectiveness in incorporating symmetry awareness during training and inference.
\textbf{3.} Across datasets ranging from Perov-5 to MPTS-52, all techniques experience a drop in performance with increasing atoms per cell. Despite this, our approach consistently surpasses the performance of other methods. In particular, our method significantly improved the RMSE metric, indicating that our equivariant diffusion approach effectively reduces the redundancy in the solution space, allowing the model to better learn the distribution characteristics of crystal data.

\subsection{Ablation Studies}
\label{sec:abla_stu}
In Table~\ref{tab:pre_abl}, we conduct an ablation study on each component of \model{}, exploring the following aspects.
\textbf{1}. To verify the necessity of lattice permutation equivariance in the generation procedure, we conduct experiments by removing the loss component of lattice permutation. Result indicates that 3.77\% decrease in match rate and 6.67\% increase in RMSE, both of these performance metrics have deteriorated. 
In addition, we also compared the performance of the Frame Average method, which is also lower than that of our proposed method. 
\textbf{2}. Without applying periodic CoM-free noising, we observe a significant deterioration in the performance metrics, with the match rate dropping from 57.59\% to 52.31\% and the RMSE increasing from 0.0510 to 0.0594. This substantial change indicates that our methodology has effectively captured the characteristics of crystalline periodic translation during training, leading to a notable impact on the evaluation metrics. \textbf{3}. To further investigate the importance of the Von Mises and Probalistic Model components in periodic translation, we conducted ablation studies on these two modules separately. Even without utilizing these two components, we observed performance improvements compared to scenarios where the noising method $\vm(\cdot)$ was not used. This indicates that our approach of modifying the noising method to ensure the score is periodic translation invariant is valid.
And we observed that removing any or all of the Von Mises and Probalistic Model components will reduce the performance of the model in terms of match rate and RMSE, indicating that both components play a positive role in performance.

\begin{table}[htp]
\vspace{-1.5em}
\setlength\tabcolsep{1.2pt}
  \centering
  \caption{Ablation studies of \model{} model on MP-20.} 
  \label{tab:pre_abl}
  \resizebox{0.95\linewidth}{!}{
  \begin{tabular}{lcc}
    \toprule
    & \multicolumn{2}{c}{Performance} \\
    \cmidrule{2-3} 
    Method  & Match rate$\uparrow$ & RMSE$\downarrow$ \\
    \midrule\midrule
    \model{} &\textbf{57.59}  &\textbf{0.0510}   \\ \midrule
    \addlinespace[-0.1pt]
     \rowcolor{black!15} \multicolumn{3}{c}{$w/o$ $Lattice$ $Permutation$ $Equivariance$} \\
    \addlinespace[-2pt]
     \midrule
    $w/o$ permutation loss  &55.42 & 0.0544  \\
    \midrule
     $w/$ Frame Average  &55.92 & 0.0578  \\
    \midrule
    \addlinespace[-0.1pt]
    \rowcolor{black!15} \multicolumn{3}{c}{$w/o$ $Periodic$ $CoM-free$ $Noising$} \\
    \addlinespace[-2pt]
    \midrule
    $w/o$ $\vm(\cdot)$ & 52.31 & 0.0594  \\
    \midrule
    \addlinespace[-0.1pt]
    \rowcolor{black!15} \multicolumn{3}{c}{$w/$ $Partial$ $Periodic$ $CoM-free$ $Noising$} \\
    \addlinespace[-2pt]
    \midrule
    $w/o$ Probalistic Model \& $w/o$ Von Mises & 54.77 & 0.0578  \\
    \midrule
    $w/$ Von Mises \& $w/o$ Probalistic Model & 57.03 & 0.0537  \\
    \midrule
    $w/$ Probalistic Model \& $w/o$ Von Mises & 56.32 & 0.0525  \\
  \bottomrule
\end{tabular}
}
\vspace{-1.5em}
\end{table}


\section{Conclusion}
In summary, we introduce \model{}, a novel equivariant diffusion generative model for Crystal Structure Prediction task. We addresses a previously unacknowledged challenge in current models: lattice permutation equivariance. During the diffusion phase, when lattice parameters undergo permutation, the fractional coordinates of atoms experience an equivariant transformation, ensuring consistency and preserving structural integrity. Furthermore, we have devised an innovative noising algorithm that meticulously preserves periodic translation equivariance throughout both the inference and training phases. Experimental results unequivocally demonstrate that \model{} outperforms existing CSP methods, achieving superior in generating high quilty structures.

\section*{Acknowledgements}

This research was jointly supported by the following project: the National Key R\&D Program of China (2021YFB0301300); the Major Program of Guangdong Basic and Applied Research (2019B030302002); Guangdong Province Special Support Program for Cultivating High-Level Talents (2021TQ06X160); Pazhou Lab Research Project (PZL2023KF0001); the Fundamental Research Funds for the Central Universities, Sun Yat-sen University (23xkjc016); the National Science and Technology Major Project under Grant 2020AAA0107300; the National Natural Science Foundation of China (No. 61925601, No. 62376276); Beijing Nova Program (No. 20230484278); Alibaba Damo Research Fund.

\section*{Impact Statement}
The development of equivariant diffusion models for CSP holds transformative potential across a broad range of scientific disciplines, including materials science, chemistry, and physics. By accurately predicting the arrangement of atoms in crystal structures, this technology can significantly expedite the discovery of novel materials. 
This, in turn, has far-reaching implications for various applications such as renewable energy, pharmaceuticals, electronics, and more, where innovative materials can lead to advancements in efficiency, efficacy, and sustainability. Moreover, the ability to predict crystal structures from their elemental compositions could reduce the need for expensive and time-consuming physical experiments, making research more accessible and accelerating the pace of innovation.

\bibliography{example_paper}
\bibliographystyle{icml2024}

\input{appendix}



\end{document}

%% file: math_commands.tex
\usepackage{amsmath,amsfonts,bm}

\def\eqref#1{equation~\ref{#1}}

\def\1{\bm{1}}

\def\rvt{{\mathbf{t}}}

\def\vzero{{\bm{0}}}
\def\vone{{\bm{1}}}

\def\va{{\bm{a}}}

\def\vc{{\bm{c}}}

\def\vf{{\bm{f}}}
\def\vg{{\bm{g}}}
\def\vh{{\bm{h}}}

\def\vk{{\bm{k}}}
\def\vl{{\bm{l}}}
\def\vm{{\bm{m}}}

\def\vs{{\bm{s}}}

\def\vx{{\bm{x}}}
\def\vy{{\bm{y}}}

\def\vepsilon{{\bm{\epsilon}}}

\def\mA{{\bm{A}}}

\def\mC{{\bm{C}}}

\def\mF{{\bm{F}}}

\def\mH{{\bm{H}}}
\def\mI{{\bm{I}}}

\def\mL{{\bm{L}}}

\def\mP{{\bm{P}}}
\def\mQ{{\bm{Q}}}

\def\mS{{\bm{S}}}

\def\mX{{\bm{X}}}

\DeclareMathAlphabet{\mathsfit}{\encodingdefault}{\sfdefault}{m}{sl}
\SetMathAlphabet{\mathsfit}{bold}{\encodingdefault}{\sfdefault}{bx}{n}

\def\gL{{\mathcal{L}}}
\def\gM{{\mathcal{M}}}
\def\gN{{\mathcal{N}}}

\def\gS{{\mathcal{S}}}

\def\gU{{\mathcal{U}}}
\def\gV{{\mathcal{V}}}

\def\sR{{\mathbb{R}}}

\def\sZ{{\mathbb{Z}}}

\newcommand{\E}{\mathbb{E}}

\newcommand{\R}{\mathbb{R}}

\DeclareMathOperator{\atan2}{atan2}


%% file: appendix.tex
\newpage


\appendix
\onecolumn

\section{Theoretical Analysis.}
\subsection{Proof of Proposition \ref{prop:lpi}}
\label{proof:pro_45}


We first introduce the following definition to describe the equivariance and invariance from the perspective of distributions.
\begin{definition}
\label{def:ei}
We call a distribution $p(x)$ is $G$-invariant if for any transformation $g$ in the group $G$, $p(g\cdot x) = p(x)$, and a conditional distribution $p(x|c)$ is G-equivariant if $p(g\cdot x|g\cdot c) = p(x|c), \forall g\in G$.
\end{definition}
We then provide the following lemma to capture the symmetry of the generation process.
\begin{lemma}[\citet{xu2021geodiff}]
\label{lm:mrkv}
Consider the generation Markov process $p(x_0) = p(x_T)\int p(x_{0:T-1}|x_t)dx_{1:T}$. If the prior distribution $p(x_T)$ is G-invariant and the Markov transitions $p(x_{t-1}|x_t), 0 < t\leq T$ are G-equivariant, the marginal distribution $p(x_0)$ is also G-invariant.
\end{lemma}

\setcounter{section}{4}
\setcounter{theorem}{0}
The proposition~\cref{prop:lpi} is rewritten and proved as follows.


\renewcommand\thesection{\Alph{section}}
\setcounter{section}{1}
\setcounter{theorem}{2}
\begin{proof}
Consider the transition probability in Eq.~(\ref{eq:ddpm1}), we have
\begin{align*}
    p(\mC_{t-1}|\mC_t,\mF_t,\mA) = \gN(\mC_{t-1}|a_t (\mC_t - b_t\hat{\vepsilon}_\mL(\mC_t,\mF_t,\mA, t)), \sigma^2_t\mI),
\end{align*}
where $a_t = \frac{1}{\sqrt{\alpha_t}}, b_t = \frac{\beta_t} {\sqrt{1 -\bar{\alpha}_t}}, \sigma^2_t=\beta_t\cdot\frac{1-\bar{\alpha}_{t-1}}{1-\bar{\alpha}_t}$ for simplicity, and 
 $\hat{\vepsilon}_\mL(\gM_t, t)$ is completed as $\hat{\vepsilon}_\mL(\mC_t,\mF_t,\mA, t)$.
 
As the denoising term $\hat{\vepsilon}_\mL(\mC_t,\mF_t,\mA, t)$ is lattice permutation equivariant, we have $\hat{\vepsilon}_\mL(\mP\mC_t,\mP\mF_t,\mA, t) = \mP\hat{\vepsilon}_\mL(\mC_t,\mF_t,\mA, t)$ for any permutation matrix $\mP\in S_3, \mP^\top\mP=\mI$.

For the variable $\mC\sim\gN(\bar{\mC},\sigma^2\mI)$, we have $\mP\mC\sim\gN(\mP\bar{\mC},\mP(\sigma^2\mI)\mP^\top)=\gN(\mP\bar{\mC},\sigma^2\mI)$. That is,
\begin{align}
\label{s1}
    \gN(\mC|\bar{\mC},\sigma^2\mI) = \gN(\mP\mC|\mP\bar{\mC},\sigma^2\mI).
\end{align}

For the transition probability $p(\mC_{t-1}|\mC_t,\mF_t,\mA)$, we have
 \begin{align*}
     p(\mP\mC_{t-1}|\mP\mC_t,\mP\mF_t,\mA) &= \gN(\mP\mC_{t-1}|a_t (\mP\mC_t - b_t\hat{\vepsilon}_\mL(\mP\mC_t,\mP\mF_t,\mA, t)), \sigma^2_t\mI) \\
      &= \gN(\mP\mC_{t-1}|a_t (\mP\mC_t - b_t\mP\hat{\vepsilon}_\mL(\mC_t,\mF_t,\mA, t)), \sigma^2_t\mI) \tag{lattice permutation equivariant $\hat{\vepsilon}_\mL$}\\
     &= \gN(\mP\mC_{t-1}|\mP\Big(a_t (\mC_t - b_t\hat{\vepsilon}_\mL(\mC_t,\mF_t,\mA, t))\Big), \sigma^2_t\mI) \\
     & =\gN(\mC_{t-1}|a_t (\mC_t - b_t\hat{\vepsilon}_\mL(\mC_t,\mF_t,\mA, t)), \sigma^2_t\mI) \tag{Eq.~(\ref{s1})}\\
     & = p(\mC_{t-1}|\mC_t,\mF_t,\mA).
 \end{align*}
 As the transition is lattice permutation equivariant and the prior distribution $\gN(0,\mI)$ is lattice permutation invariant, we prove that the the marginal distribution $p(\mC_0)$ is lattice permutation invariant based on lemma~\ref{lm:mrkv}.
\end{proof}

\subsection{Proof of Proposition \ref{prop:lpif}}
\label{proof:pro_46}

Let $\gN_w(\mu,\sigma^2\mI)$ denote the wrapped normal distribution with mean $\mu$, variance $\sigma^2$ and period 1. We first provide the following lemma.
\begin{lemma}
\label{lm:tenv}
    If the denoising term $\hat{\vepsilon}_\mF(\mC_t,\mF_t,\mA,t)$ is lattice permutation equivariant, and the transition probabilty can be formulated as $p(\mF_{t-1}|\mC_t, \mF_t, \mA) = \gN_w(\mF_{t-1}|\mF_{t} + u_t\hat{\vepsilon}_\mF(\mC_t,\mF_t,\mA,t), v^2_t\mI)$, where $u_t,v_t$ are functions of $t$, the transition is lattice permutation equivariant.
\end{lemma}

\renewcommand\thesection{\Alph{section}}
\setcounter{section}{1}
\setcounter{theorem}{2}
\begin{proof}

For the variable $\mF\sim\gN_w(\bar{\mF},v^2_t\mI)$ and $\mP\in S_3$, we have $\mP\mF\sim\gN_w(\mP\bar{\mF},\mP(v^2_t\mI)\mP^\top)=\gN_w(\mP\bar{\mF},v^2_t\mI)$. That is,
\begin{align}
\label{sf}
    \gN_w(\mF|\bar{\mF},v^2_t\mI) = \gN_w(\mP\mF|\mP\bar{\mF},v^2_t\mI).
\end{align}

For the transition probability $p(\mF_{t-1}|\mC_t,\mF_t,\mA)$, we have
 \begin{align*}
     p(\mP\mF_{t-1}|\mP\mC_t,\mP\mF_t,\mA) &= \gN_w(\mP\mF_{t-1}|\mP\mF_t + u_t\hat{\vepsilon}_\mF(\mP\mC_t,\mP\mF_t,\mA, t), v^2_t\mI) \\
      &= \gN_w(\mP\mF_{t-1}|\mP\mF_t + u_t\mP\hat{\vepsilon}_\mF(\mC_t,\mF_t,\mA, t), v^2_t\mI) \tag{lattice permutation equivariant $\hat{\vepsilon}_\mF$}\\
     &= \gN_w(\mP\mF_{t-1}|\mP\Big(\mF_t + u_t\hat{\vepsilon}_\mF(\mC_t,\mF_t,\mA, t)\Big), v^2_t\mI) \\
     & =\gN_w(\mF_{t-1}|\mF_t + u_t\hat{\vepsilon}_\mF(\mC_t,\mF_t,\mA, t), v^2_t\mI) \tag{Eq.~(\ref{sf})}\\
     & = p(\mF_{t-1}|\mC_t,\mF_t,\mA).
 \end{align*}
 \end{proof}
 
 The transition probability of the fractional coordinates during the Predictor-Corrector sampling can be formulated as
\begin{align}
    \label{eq:F_generation}
    \begin{aligned}
    p(\mF_{t-1}|\mC_t, \mF_t, \mA) &= p_P(\mF_{t-\frac{1}{2}}|\mC_t, \mF_t, \mA) p_C(\mF_{t-1}|\mC_{t-1}, \mF_{t-\frac{1}{2}}, \mA), \\
    p_P(\mF_{t-\frac{1}{2}}|\mC_t, \mF_t, \mA) &= \gN_w(\mF_{t-\frac{1}{2}}|\mF_t+(\sigma_t^2-\sigma_{t-1}^2)\hat{\vepsilon}_\mF(\mC_t, \mF_t, \mA, t), \frac{\sigma_{t-1}^2(\sigma_t^2-\sigma_{t-1}^2)}{\sigma_t^2}\mI), \\
    p_C(\mF_{t-1}|\mC_{t-1}, \mF_{t-\frac{1}{2}}, \mA) &= \gN_w(\mF_{t-\frac{1}{2}}|\mF_t+\gamma\frac{\sigma_{t-1} }{\sigma_1} \hat{\vepsilon}_\mF(\mC_{t-1}, \mF_{t-\frac{1}{2}}, \mA, t-1), 2\gamma\frac{\sigma_{t-1} }{\sigma_1}\mI),
    \end{aligned}
\end{align}
where $p_P,p_C$ are the transitions of the predictor and corrector. According to lemma~\ref{lm:tenv}, both of the transitions are lattice permutation equivariant. Therefore, the transition $p(\mF_{t-1}|\mC_t, \mF_t, \mA)$ is lattice permutation equivariant. As the prior distribution $\gU(0,1)$ is lattice permutation invariant, we finally prove that the marginal distribution $p(\mF_0)$ is lattice permutation invariant based on lemma~\ref{lm:mrkv}.


\subsection{Proof of Periodic CoM-free Nosing}
\label{proof:p_com}

We prove the periodic CoM-free function $\vm(\vepsilon)$ constructed by Eq (\ref{eq:pcom}) is periodic translation invariance as Eq (\ref{eq:m_invariant}) describes in this section. Since $\vm(\cdot)$ can be treated as operating on each row of $\vepsilon\in\R^{3\times n}$ independently, without loss of generality, we use $\vepsilon=[\epsilon_1, \epsilon_2, \ldots, \epsilon_n]$ to represent any row of the origin $\vepsilon$ for simplification. Then we can rewrite $\vm$ as:
\begin{align*}
\left\{
\begin{aligned}
\vm(\vepsilon) &= w\big(\vepsilon-\frac{\atan2 \left(\bar{\vy}(\vepsilon), \bar{\vx}(\vepsilon)\right)}{2\pi}\big),\\
\bar{\vy}(\vepsilon) &=\frac1n\sum_{i=0}^n\sin{(2\pi\epsilon_i)},\\
\bar{\vx}(\vepsilon) &=\frac1n\sum_{i=0}^n\cos{(2\pi \epsilon_i)},
\end{aligned}
\right.
\end{align*}
We next prove that $\vm(\vepsilon)=\vm(\vepsilon+r)$ for any $r\in\R$. 
\begin{proof}
We can rewrite $\vm(\vepsilon)=\vm(\vepsilon+r)$:
\begin{align*}
    w\big(\vepsilon-\frac{\atan2 \left(\bar{\vy}(\vepsilon), \bar{\vx}(\vepsilon)\right)}{2\pi}\big) &=  w\big(\vepsilon+r-\frac{\atan2 \left(\bar{\vy}(\vepsilon+r), \bar{\vx}(\vepsilon+r)\right)}{2\pi}\big)\\
    \vepsilon-\frac{\atan2 \left(\bar{\vy}(\vepsilon), \bar{\vx}(\vepsilon)\right)}{2\pi}&=  \vepsilon+r-\frac{\atan2 \left(\bar{\vy}(\vepsilon+r), \bar{\vx}(\vepsilon+r)\right)}{2\pi} + d ,\\
\end{align*}
where $d$ denotes any integer. We can redefine that $r=w(r)\in[0,1)$ because  $\bar{\vy}(\vepsilon+r)=\bar{\vy}(\vepsilon+w(r))$ and $\bar{\vx}(\vepsilon+r)=\bar{\vx}(\vepsilon+w(r))$ by the periodicity of trigonometric functions and we can merge  integer part of the origin $r$ into $d$ since $d$ can be any integer. Further simplification leads to:
\begin{align*}
    \frac{\atan2 \left(\bar{\vy}(\vepsilon), \bar{\vx}(\vepsilon)\right)}{2\pi} + r + d &= \frac{\atan2 \left(\bar{\vy}(\vepsilon+r), \bar{\vx}(\vepsilon+r)\right)}{2\pi},\\
    \atan2 \left(\bar{\vy}(\vepsilon), \bar{\vx}(\vepsilon)\right) + 2\pi\cdot r + 2\pi\cdot d &= \atan2 \left(\bar{\vy}(\vepsilon+r), \bar{\vx}(\vepsilon+r)\right),\\
    \tan\big(\atan2 \left(\bar{\vy}(\vepsilon), \bar{\vx}(\vepsilon)\right) + 2\pi\cdot r + 2\pi\cdot d\big) &= \frac{\bar{\vy}(\vepsilon+r)}{\bar{\vx}(\vepsilon+r)} \tag{$\tan(\cdot)$ for both sides} ,\\
    \tan\big(\atan2 \left(\bar{\vy}(\vepsilon), \bar{\vx}(\vepsilon)\right) + 2\pi\cdot r) &= \frac{\bar{\vy}(\vepsilon+r)}{\bar{\vx}(\vepsilon+r)} ,\\
    \frac{\frac{\bar{\vy}(\vepsilon)}{\bar{\vx}(\vepsilon)}+\tan(2\pi\cdot r)}{1-\frac{\bar{\vy}(\vepsilon)}{\bar{\vx}(\vepsilon)}\tan(2\pi\cdot r)} &= \frac{\bar{\vy}(\vepsilon+r)}{\bar{\vx}(\vepsilon+r)} \tag{$\tan(\cdot)$ addition formula} ,\\
    \frac
    {\bar{\vy}(\vepsilon)\cos(2\pi\cdot r)+\bar{\vx}(\vepsilon)\sin(2\pi\cdot r)}
    {\bar{\vx}(\vepsilon)\cos(2\pi\cdot r)-\bar{\vy}(\vepsilon)\sin(2\pi\cdot r)} &= \frac{\bar{\vy}(\vepsilon+r)}{\bar{\vx}(\vepsilon+r)},
\end{align*}
Then we can prove that:
\begin{align}
\text{the right-hand side} &= 
    \frac
    {\frac1n\sum_{i=0}^n\sin{(2\pi\epsilon_i+2\pi\cdot r)}}
    {\frac1n\sum_{i=0}^n\cos{(2\pi \epsilon_i+2\pi\cdot r)}},\nonumber \\
    &= 
    \frac
    {\frac1n\sum_{i=0}^n\big(\sin{(2\pi\epsilon_i)}\cos{2\pi\cdot r}+\cos{(2\pi\epsilon_i)}\sin{(2\pi\cdot r)}\big)}
    {\frac1n\sum_{i=0}^n\big(\cos{(2\pi \epsilon_i)}\cos{(2\pi\cdot r)}-\sin{(2\pi \epsilon_i)}\sin{(2\pi\cdot r)}\big)}\tag{$\sin(\cdot)$ and $\cos(\cdot)$ addition formula}\\
    \label{eq:xyprop}
    &= 
    \frac
    {\bar{\vy}(\vepsilon)\cos(2\pi\cdot r)+\bar{\vx}(\vepsilon)\sin(2\pi\cdot r)}
    {\bar{\vx}(\vepsilon)\cos(2\pi\cdot r)-\bar{\vy}(\vepsilon)\sin(2\pi\cdot r)} \\
    &= 
    \text{the left-hand side}\nonumber
\end{align}

We finally prove that $\vm(\vepsilon)$ is periodic translation invariance. 

\end{proof}

\subsection{Proof of Proposition \ref{prop:ptiexp}}
\label{apd:mgrad}

Since $\nabla_\vepsilon\big( \vm({\vepsilon})[:,i]\big)$ can be treated as operating on each row of $\vepsilon\in\R^{3\times n}$ independently, without loss of generality, we use $\vepsilon=[\epsilon_1, \epsilon_2, \ldots, \epsilon_n]$ to represent any row of the origin $\vepsilon$, and use $\vm_i({\vepsilon})$ to represent the origin $\vm({\vepsilon})[:,i]$ for simplification. Then we can rewrite $\vm({\vepsilon})[:,i]$ as:
\begin{align}
\label{eq:simple_m}
\left\{
\begin{aligned}
\vm_i(\vepsilon) &= w\big(\epsilon_i-\frac{\atan2 \left(\bar{\vy}(\vepsilon), \bar{\vx}(\vepsilon)\right)}{2\pi}\big),\\
\bar{\vy}(\vepsilon) &=\frac1n\sum_{j=0}^n\sin{(2\pi\epsilon_j)},\\
\bar{\vx}(\vepsilon) &=\frac1n\sum_{j=0}^n\cos{(2\pi \epsilon_j)},
\end{aligned}
\right.
\end{align}

Our target is to prove that $\nabla_\vepsilon \vm_i({\vepsilon})$ is periodic translation invariance. 

We first get the formula of $\nabla_\vepsilon \vm_i({\vepsilon})$:
\begin{align*}
\frac{\partial \vm_i}{\partial \epsilon_j} &= 
\left\{
\begin{aligned}
 - \frac{1}{2\pi}\left(\frac{\bar{x}}{\bar{x}^2 + \bar{y}^2} \cdot \frac{\partial \bar{y}}{\partial \epsilon_j} - \frac{\bar{y}}{\bar{x}^2 + \bar{y}^2} \cdot \frac{\partial \bar{x}}{\partial \epsilon_j}\right),\text{ if } & i\neq j, \\
  1 - \frac{1}{2\pi}\left(\frac{\bar{x}}{\bar{x}^2 + \bar{y}^2} \cdot \frac{\partial \bar{y}}{\partial \epsilon_j} - \frac{\bar{y}}{\bar{x}^2 + \bar{y}^2} \cdot \frac{\partial \bar{x}}{\partial \epsilon_j}\right),\text{ if } & i = j,
\end{aligned}
\right.
\end{align*}
where
\begin{align*}
\bar{y}&=\bar{\vy}(\vepsilon),\\
\bar{x}&=\bar{\vx}(\vepsilon),\\
\frac{\partial \bar{y}}{\partial \epsilon_j} &= \frac{1}{n} \cdot 2\pi \cos(2\pi\epsilon_j), \\
\frac{\partial \bar{x}}{\partial \epsilon_j} &= -\frac{1}{n} \cdot 2\pi \sin(2\pi\epsilon_j).
\end{align*}
Substituting these partial derivatives, we obtain:
\begin{align}
\label{eq:mgrad}
\frac{\partial \vm_i}{\partial \epsilon_j} &= 
\left\{
\begin{aligned}
- \frac{1}{n(\bar{x}^2 + \bar{y}^2)} \left(\bar{x} \cos(2\pi\epsilon_j) + \bar{y} \sin(2\pi\epsilon_j)\right),\text{ if } & i\neq j, \\
1 - \frac{1}{n(\bar{x}^2 + \bar{y}^2)} \left(\bar{x} \cos(2\pi\epsilon_j) + \bar{y} \sin(2\pi\epsilon_j)\right),\text{ if } & i = j,
\end{aligned}
\right.
\end{align}
where $\bar{y}=\bar{\vy}(\vepsilon)$ and $\bar{x}=\bar{\vx}(\vepsilon)$. Thus, the gradient $\nabla_\vepsilon \vm_i({\vepsilon})$ is a vector with its $j^{th}$ component given by Eq.(\ref{eq:mgrad}).

We next prove $\nabla_\vepsilon \vm_i({\vepsilon})=\nabla_{\vepsilon+r} \vm_i({\vepsilon+r})$. 

\begin{proof}
By Eq.(\ref{eq:mgrad}), we can rewrite $\nabla_\vepsilon \vm_i({\vepsilon})=\nabla_{\vepsilon+r} \vm_i({\vepsilon+r})$ as:
\begin{align}
\label{eq:rewrite_grad}
\frac
{\bar{\vx}(\vepsilon) \cos(2\pi\epsilon_j) + \bar{\vy}(\vepsilon) \sin(2\pi\epsilon_j)}
{\bar{\vx}^2(\vepsilon) + \bar{\vy}^2(\vepsilon)} =
\frac
{\bar{\vx}(\vepsilon+2\pi r) \cos(2\pi\epsilon_j+2\pi r) + \bar{\vy}(\vepsilon+2\pi r) \sin(2\pi\epsilon_j+2\pi r)}
{\bar{\vx}^2(\vepsilon+2\pi r) + \bar{\vy}^2(\vepsilon+2\pi r)}
\end{align}
Referring to Eq.(\ref{eq:xyprop}), we have:
\begin{align*}
\bar{\vy}(\vepsilon+r) &= \bar{\vy}(\vepsilon)\cos(2\pi\cdot r)+\bar{\vx}(\vepsilon)\sin(2\pi\cdot r), \\
\bar{\vx}(\vepsilon+r) &= \bar{\vx}(\vepsilon)\cos(2\pi\cdot r)-\bar{\vy}(\vepsilon)\sin(2\pi\cdot r).
\end{align*}

The numerator of the right-hand side of Eq.(\ref{eq:rewrite_grad}) can be expressed as:
\begin{align*}
&\bar{x}(\vepsilon + r) \cos(2\pi\epsilon_j + 2\pi r) + \bar{y}(\vepsilon + r) \sin(2\pi\epsilon_j + 2\pi r) \\
&= [\bar{x}(\vepsilon)\cos(2\pi r) - \bar{y}(\vepsilon)\sin(2\pi r)] [\cos(2\pi\epsilon_j)\cos(2\pi r) - \sin(2\pi\epsilon_j)\sin(2\pi r)] \\
&\quad + [\bar{y}(\vepsilon)\cos(2\pi r) + \bar{x}(\vepsilon)\sin(2\pi r)] [\sin(2\pi\epsilon_j)\cos(2\pi r) + \cos(2\pi\epsilon_j)\sin(2\pi r)] \\
&= \bar{x}(\vepsilon)\cos(2\pi\epsilon_j) + \bar{y}(\vepsilon)\sin(2\pi\epsilon_j),
\end{align*}
where the terms involving $r$ cancel out due to trigonometric identities. 

The denominator remains invariant under the transformation due to the Pythagorean identity:
\begin{align*}
\bar{x}^2(\vepsilon + r) + \bar{y}^2(\vepsilon + r) &= [\bar{x}(\vepsilon)\cos(2\pi r) - \bar{y}(\vepsilon)\sin(2\pi r)]^2 + [\bar{y}(\vepsilon)\cos(2\pi r) + \bar{x}(\vepsilon)\sin(2\pi r)]^2 \\
&= \bar{x}^2(\vepsilon) + \bar{y}^2(\vepsilon).
\end{align*}

Therefore, the given statement is proven:
\begin{align*}
\frac
{\bar{x}(\vepsilon) \cos(2\pi\epsilon_j) + \bar{y}(\vepsilon) \sin(2\pi\epsilon_j)}
{\bar{x}^2(\vepsilon) + \bar{y}^2(\vepsilon)} =
\frac
{\bar{x}(\vepsilon + r) \cos(2\pi\epsilon_j + 2\pi r) + \bar{y}(\vepsilon + r) \sin(2\pi\epsilon_j + 2\pi r)}
{\bar{x}^2(\vepsilon + r) + \bar{y}^2(\vepsilon + r)}.
\end{align*}

We finally prove that $\nabla_\vepsilon \vm_i({\vepsilon})$ is periodic translation invariance, which is equivalent to the statement that $\nabla_\vepsilon\big( \vm({\vepsilon})[:,i]\big)$ is periodic translation invariance. 
\end{proof}

We next prove that $\nabla_{\mF_t}\log q(\mF_t)$ is periodic translation invariance. We can derive it as:
\begin{align*}
\label{eq:score_before}
 \nabla_{\mF_t}\log q(\mF_t) &=\sum_{i=0}^n \big( \nabla_{\bar{\vf_i}} \log q(\bar{\vf_i})\big)\vone^\top  \odot\nabla_{\mF_t} \bar{\vf_i},\\
 &=\sum_{i=0}^n
 \big( \nabla_{\bar{\vf_i}} \log q(\bar{\vf_i})\big)\vone^\top
 \odot\nabla_\vepsilon\big( \vm({\vepsilon})[:,i]\big),\\
\end{align*}
Since $\nabla_{\bar{\vf_i}} \log q(\bar{\vf_i})$ is periodic translation invariance, and $\nabla_\vepsilon\big( \vm({\vepsilon})[:,i]\big)$ is also periodic translation invariance from the above proof, we can easily get $\nabla_{\mF_t}\log q(\mF_t)$ is periodic translation invariance. 

\section{Methods}
\subsection{Symmetries of crystal structure distribution}
\label{method:symme}
We represent the fractional coordinates on a lattice base of crystal as the points on the circle in Figure~\ref{fig:e3_vari} (e) (f). If the points rotate alongside the circle, it means that the fractional coordinates on the base undergo a periodic translation. Specifically, the geometry of $2 \pi \cdot f_i$ alongside the circle is equivalent to $2 \pi w(f_i + d)$. Periodic translation invariance can be explained that any rotation on the circle does not change geometry of the angle distribution.

\subsection{Diffusion on Lattice Parameters}
\label{ddpm_L}
In DDPM models~\citep{ho2020denoising}, lattice parameters typically range from \([0, +\infty)\) for lengths and \((0, \pi)\) for angles. However, DDPMs diffuse within the \((- \infty, +\infty)\) interval, potentially generating unreasonable lattice parameters during diffusion generation. To ensure generated lattice parameters are always reasonable, we apply a logarithmic transformation to lengths, as the function $\log$ maps \((0, +\infty)\) to \((- \infty, +\infty)\), perfectly aligning with our requirement. Thus, we generate $\log \vl$ instead of $\vl$, and convert it back using \(e^{\log \vl} = \vl\), ensuring lengths \(\vl\) are always positive. For angles, we process them with \(\tan(\boldsymbol{\phi} - \pi/2)\), which also maps the desired \((0, \pi)\) to \((- \infty, +\infty)\). Upon generating values for \(\tan(\boldsymbol{\phi} - \pi/2)\), we retrieve angles \(\boldsymbol{\phi}\) in the \((0, \pi)\) range through \(\arctan(\tan(\boldsymbol{\phi} - \pi/2) + \pi/2) = \boldsymbol{\phi}\).

\subsection{Frame Average Method}
\label{appen:fa}

Frame Average Method encodes the lattice permutation equivariance to the neural network. On the context of lattice permutation group, a frame is defined as one specific order of lattice vectors. By applying a permutation matrix to both the lattice and its fractional coordinates, we are able to transform the structure into an equivalent frame, as we described in Definition~\ref{prop:lpi} of our paper.

Specifically, we adjusted the neural network formula to: 
\begin{align*}
\phi_{FA}(\mX) = \frac{1}{6}\sum_{\mP\in S_3}\mP^{-1}\phi(\mP \mX)
\end{align*}
where $\phi$ is the neural network model in Section~\ref{sec:model}, $\mX$ is a $3 \times N$ matrix comprising three lattices and their respective fractional coordinates, and $\mP$ represents the permutation matrices for the lattices. 

\section{Implementation Details.}

\subsection{Von Mises Distribution Simulation}
\label{sec:von_mises}




The Von Mises distribution\cite{gatto2007generalized}, often referred to as the 'circular normal' distribution, is a probability distribution used for modeling angular or directional data. It is flexible and efficient to handle periodic and directional characteristics. Let $\gV(\mu,\kappa)$ denote the Von Mises distribution with mean direction $\mu$, concentration parameter $\kappa$ and period 1. The probability density function (PDF) of $\gV(\mu,\kappa)$ is defined as: 
\begin{align*}
    \gV(x; \mu, \kappa) = \frac{e^{\kappa \cos(2\pi x - \mu)}}{I_0(\kappa)},
\end{align*}
where $\mu$ is the mean direction of the distribution, and $\kappa$ is the concentration parameter, indicating the level of concentration around the mean direction. The function $I_0(\kappa)$ is the modified Bessel function of order zero, which normalizes the distribution. 

As detailed in Section \ref{generation for frac}, we employ Von Mises distribution to simulate $p(\bar{\vepsilon}_i)$ where $\bar{\vepsilon}=\vm(\vepsilon)$ by Eq.(\ref{eq:pcom}), $\vepsilon\in\R^{3\times n}$ and $\vepsilon\sim\gN_w(0, \sigma_t^2\mI)$ and $\bar{\vepsilon}_i$ is the i-coloumn of $\bar{\vepsilon}$. Similar to the analysis in Appendix.\ref{apd:mgrad}, we can simplify the question as: using $\gV(\mu,\kappa)$ to simulate $p(\bar{\epsilon}_i)$ where $\bar{\epsilon}_i=\vm_i(\vepsilon)$ by Eq.(\ref{eq:simple_m}), $\vepsilon=[\epsilon_1, \epsilon_2, \cdots \epsilon_n]$, $i\in\gU(1, n)$ and $\vepsilon\sim\gN_w(0, \sigma_t^2\mI)$. Since the mean of the $\vepsilon\sim\gN_w(0, \sigma_t^2\mI)$ is 0 and the function $\vm_i(\vepsilon)$  intuitively moves the elements of $\vepsilon$ as a whole closer to 0, we set the mean direction of $\gV(\mu,\kappa)$ to 0 empirically. As a result, the key of simulation is to estimate the concentration parameter $\kappa$. We denotes $\gV(0,\kappa(n, \sigma_t))$ as the target distribution since $\kappa$ is relative to the size of $\vepsilon$ i.e. $n$ and the variance of $\gN_w(0, \sigma_t^2\mI)$ i.e. $\sigma_t$. 

We employ the Monte Carlo method to estimate \(\kappa(n, \sigma_t)\). For each specified \(n\) and \(\sigma_t\), the procedure initiates by generating samples from \(\vepsilon\sim\gN_w(0, \sigma_t^2\mI)\), which are then transformed into \(\bar{\epsilon}_i\) following the methodology outlined above. These transformed points are then used to compute their respective probability values according to \(\gV(0,\kappa(n, \sigma_t))\), where \(\kappa(n, \sigma_t)\) is initially set to an arbitrary value. The negative log-likelihood of these probabilities serves as the loss function. To refine the estimation and ascertain the optimal \(\kappa(n, \sigma_t)\) value, we utilize the minimize function from the SciPy library, ensuring an effective and precise optimization tailored to each \(n\) and \(\sigma_t\) configuration.

We have obtained the approximate probability density function for each element of $\bar{\vepsilon}$. As each element is considered to be independently and identically distributed, the calculation of the score for $\bar{\vepsilon}$ involves deriving the corresponding score for each individual element, i.e.$\nabla_{\bar{\epsilon}_i} \log \gV(\bar{\epsilon}_i | 0,\kappa(n, \sigma_t))$.
To derive \(\nabla_{\bar{\epsilon}_i} \log \gV(\bar{\epsilon}_i | 0,\kappa(n, \sigma_t))\), we first consider the logarithm of $\gV(\bar{\epsilon}_i; \mu, \kappa)$:

\begin{align*}
\log \gV(\bar{\epsilon}_i; \mu, \kappa) &= \log\left(\frac{e^{\kappa \cos(2\pi \bar{\epsilon}_i - \mu)}}{I_0(\kappa)}\right) \\
&= \kappa \cos(2\pi \bar{\epsilon}_i - \mu) - \log I_0(\kappa)
\end{align*}

Now, taking the gradient with respect to \(\bar{\epsilon}_i\), we get:

\begin{equation}
\label{eq:simple_v}
\begin{aligned}
\nabla_{\bar{\epsilon}_i} \log \gV(\bar{\epsilon}_i | 0,\kappa(n, \sigma_t)) &= \frac{d}{d\bar{\epsilon}_i} \left(\kappa(n, \sigma_t) \cos(2\pi \bar{\epsilon}_i) - \log I_0(\kappa(n, \sigma_t))\right) \\
&= -2\pi \kappa(n, \sigma_t) \sin(2\pi \bar{\epsilon}_i)
\end{aligned}
\end{equation}

To sum up, we have:

\begin{align*}
\nabla_{\bar{\vepsilon}_i} \log p(\bar{\vepsilon}_i) &\approx
\nabla_{\bar{\vepsilon}_i} \log \gV(\bar{\vepsilon}_i | 0,\kappa(n, \sigma_t))\\
&= -2\pi \kappa(n, \sigma_t) \sin(2\pi \bar{\vepsilon}_i)
\end{align*}

\subsection{Probabilistic Modeling Process}
\label{probab_model_pro}

We simplify Eq.(\ref{eq:score_after}) to improve the efficiency of calculation. Since $\nabla_{\mF_t}\log q(\mF_t)$ can be treated as operating on each row of $\bar{\vepsilon}$ independently, without loss of generality, we use $\bar{\vepsilon}=[\bar{\epsilon}_1, \bar{\epsilon}_2, \ldots, \bar{\epsilon}_n]$ to represent any row of the origin $\bar{\vepsilon}$ and use $\vs$ to represent the corresponding row of $\nabla_{\mF_t}\log q(\mF_t)$ for simplification, while using $\bar{\vs}=[\bar{s}_1, \bar{s}_2,\ldots,\bar{s}_n]$ to represent the corresponding row of $\nabla_{\bar{\mF}}\log q(\bar{\mF})$. Then we have:

\begin{align*}
 \vs &= \sum_{i=0}^n\big(\bar{s}_i \cdot\nabla_{\bar{\vepsilon}}  \vm_i({\bar{\vepsilon}})\big),
\end{align*}
where $\nabla_{{\bar{\epsilon}}_i}\log p({\bar{\epsilon}}_i)$ can be obtained by Eq.(\ref{eq:simple_v}) and $\nabla_{\bar{\vepsilon}}  \vm_i({\bar{\vepsilon}})$ can be obtained by Eq.(\ref{eq:mgrad}). Further expanding:
\begin{align*}
\vs &= 
 \bar{s}_1\cdot[1+g_1, g_2,\ldots, g_n]+\bar{s}_2\cdot[g_1, 1+g_2,\ldots, g_n]+\ldots\\
 &\quad\quad\bar{s}_n\cdot[g_1, g_2,\ldots, 1+g_n]
 ,\\
 &=
 \Bigg[\bar{s}_1+g_1\cdot\big(\sum_{i=0}^n \bar{s}_i\big), \quad \ldots, \quad \bar{s}_n+g_n\cdot\big(\sum_{i=0}^n\bar{s}_i\big)\Bigg],\\
  &=
\bar{\vs} + \big(\sum_{i=0}^n\bar{s}_i\big)\cdot\vg(\bar{\vepsilon}),
\end{align*}
where 
\begin{align*}
\left\{
\begin{aligned}
\vg(\bar{\vepsilon}) &= [g_1, g_2, \ldots, g_n], \\
&= - \frac{\bar{x}\cos(2\pi\bar{\vepsilon}) + \bar{y} \sin(2\pi\bar{\vepsilon})}{n(\bar{x}^2 + \bar{y}^2)} \\
\bar{x} &= \frac1n\sum_{j=0}^n\sin{(2\pi\bar{\epsilon}_j)},\\
\bar{y} &= \frac1n\sum_{j=0}^n\cos{(2\pi\bar{\epsilon}_j)}.
\end{aligned}
\right.
\end{align*}

Consequently, we have successfully formulated a more parallelization-friendly expression of $s$. By extrapolating this to the initial context where $\bar{\vepsilon}=[\bar{\vepsilon}_1, \bar{\vepsilon}_2, \ldots, \bar{\vepsilon}_n] \in \R^{3 \times n}$, the ultimate expression is thus deduced:
\begin{align}
\label{eq:final_score}
\nabla_{\mF_t}\log q(\mF_t) &=
\nabla_{\bar{\mF}}\log q(\bar{\mF}) + \big(\sum_{i=0}^n \bar{\vs}_i\vone^\top\odot\vg(\bar{\vepsilon}\big),
\end{align}
where 
\begin{align}
\label{eq:final_g}
\left\{
\begin{aligned}
\vg(\bar{\vepsilon})
&= - \frac{1}{n(\bar{\vx}^2 + \bar{\vy}^2)}\odot \big(\bar{\vx}\odot\cos(2\pi\bar{\vepsilon}) + \bar{\vy}\odot\sin(2\pi\bar{\vepsilon}) \big) \\
\bar{\vx} &= \big(\frac1n\sum_{j=0}^n\sin{(2\pi\bar{\vepsilon}_j)}\big)\vone^\top,\\
\bar{\vy} &= \big(\frac1n\sum_{j=0}^n\cos{(2\pi\bar{\vepsilon}_j)}\big)\vone^\top.
\end{aligned}
\right.
\end{align}

\subsection{Algorithms for Training and Sampling}
\label{apd:algo}
Algorithm~\ref{alg:train} provides a comprehensive overview of the forward diffusion process and the training procedure for the denoising model $\phi$, while Algorithm~\ref{alg:gen} elucidates the backward sampling process. These algorithms can effectively preserve symmetries if $\phi$ is meticulously designed. It is worth mentioning that we employ the predictor-corrector sampler~\citep{song2020score} to sample $\mF_{0}$ in Algorithm~\ref{alg:gen}, where Line 8 denotes the predictor, and Lines 11-12 correspond to the corrector. The $\vm(\cdot)$ denotes the function in Eq.(\ref{eq:pcom}). The $w(\cdot)$ denotes the truncation function. The $\vg(\cdot)$ denotes the function in Eq.(\ref{eq:final_g}). 

\begin{algorithm}[h]
\small
\caption{Training Procedure of \model{}}\label{alg:train}
\begin{algorithmic}[1]
\STATE \textbf{Input:} lattice parameters $\mC_0$, atom types $\mA$, fractional coordinates $\mF_0$, denoising model $\phi$, and the number of sampling steps $T$.
\STATE Sample $\vepsilon_\mL\sim\gN(\vzero,\mI)$,$\vepsilon_\mF\sim\gN(\vzero,\mI)$,$\mP \sim \mathcal{U} (\mathrm{S}_3)$ and $t\sim \gU(1,T)$.
\STATE $\bar{\vepsilon} \gets \vm(w(\sigma_t\vepsilon_F))$
\STATE $\mC_t \gets \sqrt{\bar{\alpha}_t}\mC_0 + \sqrt{1 - \bar{\alpha}_t} \vepsilon_\mL$
\STATE $\mF_t \gets w(\mF_0 + \bar{\vepsilon})$
\STATE $\hat{\vepsilon}_\mL, s_\theta, F_\theta \gets \phi(\mC_t, \mF_t, \mA, t)$
\STATE $\hat{\vepsilon}_\mL', s_\theta', F_\theta' \gets \phi(\mP\mC_t, \mP\mF_t, \mA, t)$
\STATE $\gL_\mC \gets \|\vepsilon_\mL - \hat{\vepsilon}_\mL\|_2^2$
\STATE $\gL_s \gets  \| (-2\pi\kappa(n, \sigma_t)\sin(2\pi\bar{\vepsilon})) - s_\theta\|_2^2$
\STATE $\gL_F \gets  \| \bar{\vepsilon} - F_\theta\|_2^2$
\STATE $\gL_{p\mC} \gets  \| \mP\hat{\vepsilon}_\mL - \hat{\vepsilon}_\mL'|_2^2$
\STATE $\gL_{ps} \gets  \| \mP s_\theta - s_\theta'|_2^2$
\STATE $\gL_{pF} \gets  \| \mP F_\theta - F_\theta'|_2^2$
\STATE Minimize $\gL_\mC+\gL_s+\gL_F+\gL_{p\mC}+\gL_{ps}+\gL_{pF}$
\end{algorithmic}
\end{algorithm}

\vskip -0.1in
\begin{algorithm}[h]
\small
\caption{Sampling Procedure of \model{}}\label{alg:gen}
\begin{algorithmic}[1]
\STATE \textbf{Input:} atom types $\mA$, denoising model $\phi$, number of sampling steps $T$, step size of Langevin dynamics $\gamma$.
\STATE Sample $\mC_T\sim\gN(\vzero,\mI)$,$\mF_T\sim \gU(0,1)$.
\FOR{$t \gets T,\cdots, 1$}
    \STATE{Sample $\vepsilon_\mL,\vepsilon_\mF, \vepsilon'_\mF\sim\gN(\vzero,\mI)$}
    \STATE{$\hat{\vepsilon}_\mL, s_\theta, F_\theta \gets \phi(\mC_t, \mF_t, \mA, t)$.}
    \STATE{$\mC_{t-1} \gets \frac{1}{\sqrt{\alpha_t}} (\mC_t - \frac{\beta_t} {\sqrt{1 -\bar{\alpha}_t}}\hat{\vepsilon}_\mL)  + \sqrt{\beta_t\cdot\frac{1-\bar{\alpha}_{t-1}}{1-\bar{\alpha}_t}}\vepsilon_\mL$.} 
    \STATE{
    $
    \hat{\vepsilon}_\mF = s_\theta + (\sum_{i=0}^n s_\theta[:,i])\vone^\top\odot\vg(F_\theta)
    $
    }
    \STATE{$\mF_{t-\frac{1}{2}} \gets w(\mF_t + (\sigma_t^2-\sigma_{t-1}^2)\hat{\vepsilon}_\mF + \frac{\sigma_{t-1}\sqrt{\sigma_t^2-\sigma_{t-1}^2}}{\sigma_t} \vepsilon_\mF)$}
    \STATE{$\_, s_\theta, F_\theta \gets \phi(\mC_{t-1}, \mF_{t-\frac{1}{2}}, \mA, t - 1)$.}
    \STATE{
    $
    \hat{\vepsilon}_\mF = s_\theta + (\sum_{i=0}^n s_\theta[:,i])\vone^\top\odot\vg(F_\theta)
    $
    }
    \STATE{$d_t\gets \gamma\sigma_{t-1} / \sigma_1$}
    \STATE{$\mF_{t-1} \gets w(\mF_{t-\frac{1}{2}} + d_t \hat{\vepsilon}_\mF + \sqrt{2 d_t}\vepsilon'_\mF)$.}
\ENDFOR
\STATE \textbf{Return} $\mC_0, \mF_0$.
\end{algorithmic}
\end{algorithm}

\subsection{Hyper-parameters and Training Details.}
\label{hyper_details}

For our \model{}, we employ a 4-layer setting with 256 hidden states for Perov-5 and a 6-layer setting with 512 hidden states for other datasets. The dimension of the Fourier embedding is set to $k=256$. We utilize the cosine scheduler with $s=0.008$ to regulate the variance of the DDPM process on $\mC_t$, and an exponential scheduler with $\sigma_1=0.005,\sigma_T=0.5$ to control the noise scale of the score matching process on $\mF_t$. The diffusion step is set to $T=1000$. Our model undergoes training for 3500, 4000, 1000, and 1000 epochs respectively for Perov-5, Carbon-24, MP-20, and MPTS-52 using the same optimizer and learning rate scheduler as CDVAE. For Langevin dynamics' step size $\gamma$, we apply values of $\gamma=5\times10^{-7}$ for Perov-5, $\gamma=5\times10^{-6}$ for MP-20, $\gamma=1\times10^{-5}$ for MPTS-52; while for ab initio generation in Carbon-24 case we use $\gamma=1\times10^{-5}$. All models are trained on one Nvidia A800 GPU.

\section{Learning Curves of Different Variants.}
\label{learn_curves}

\begin{figure}
    \centering
    \includegraphics[width=0.8\textwidth]{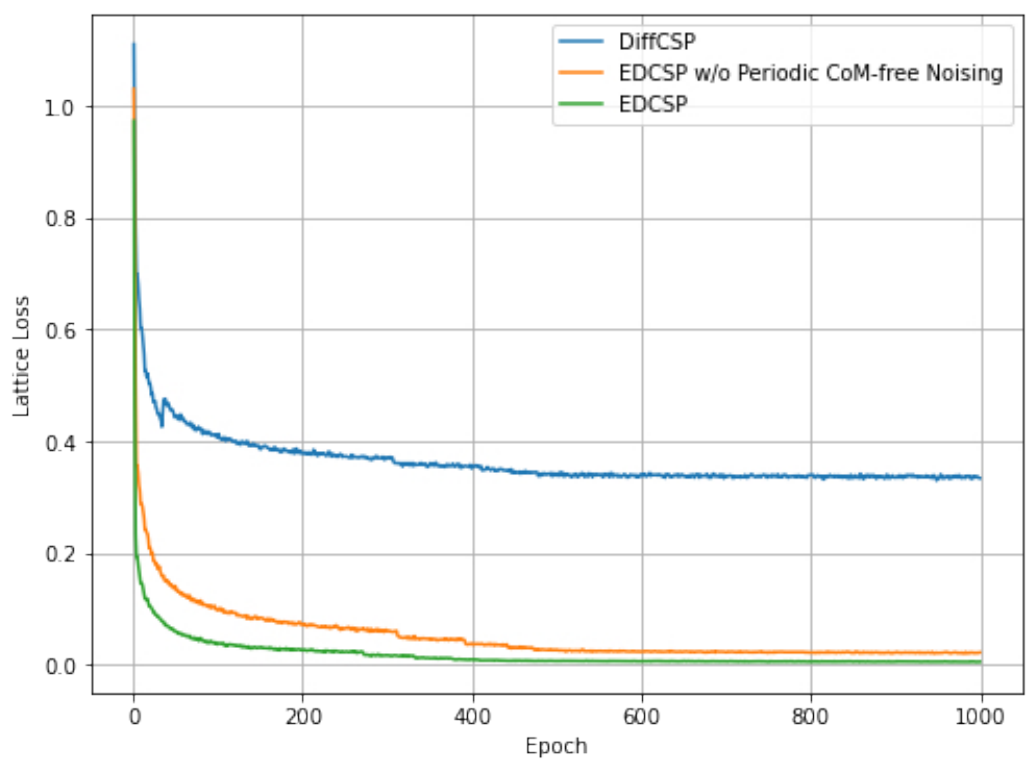}
    \caption{Learning curves of lattice loss.}
    \label{fig:lr_curve_L}
\end{figure}

\begin{figure}
    \centering
    \includegraphics[width=0.8\textwidth]{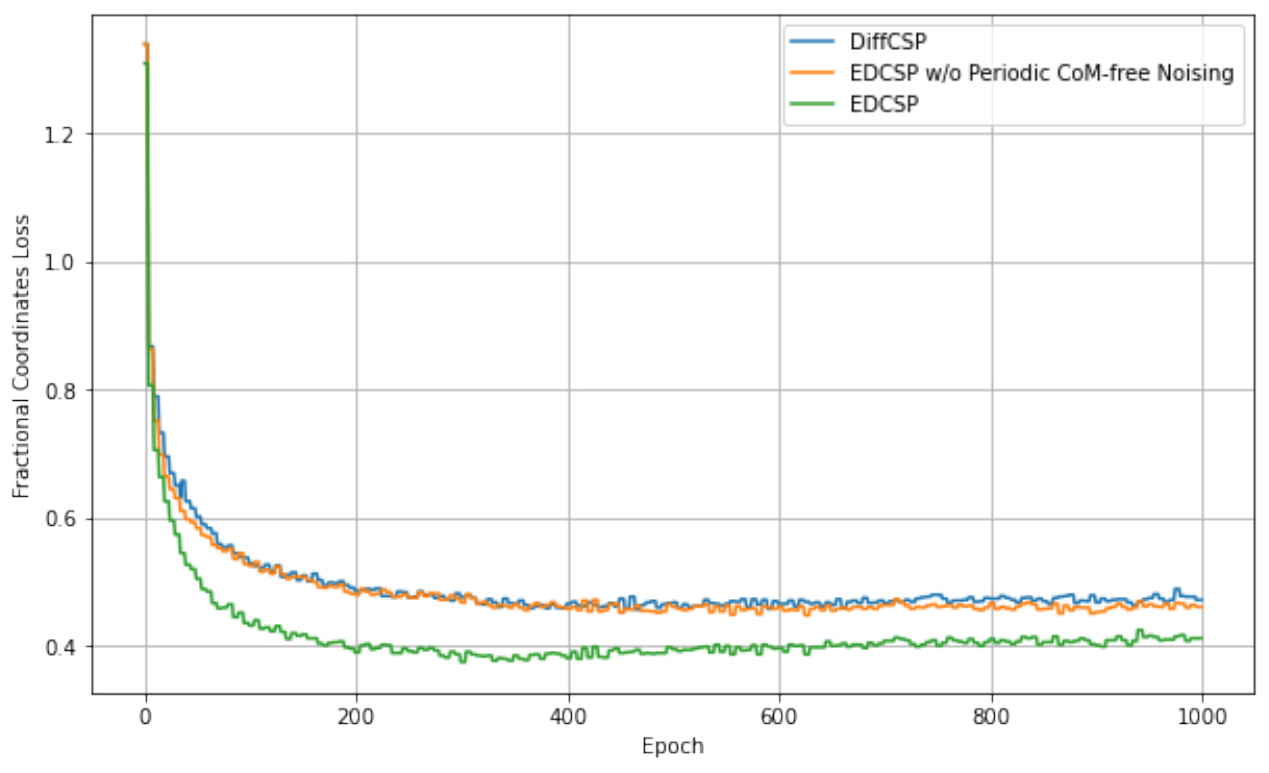}
    \caption{Leanring curves of fractional coordinates loss.}
    \label{fig:lr_curve_F}
\end{figure}

We plot the curves of training loss of different variants proposed in Figure~\ref{fig:lr_curve_L} and~\ref{fig:lr_curve_F}.

\section{Ab initio Structure Generation}
\label{sec:ab_initio_gen}
\textbf{Dataset.}
We conduct experiments on \textbf{Perov-5}, \textbf{Carbon-24} and \textbf{MP-20} dataset. Notably,
\textbf{Carbon-24}~\citep{carbon2020data} encompasses 10,153 carbon materials, each containing 6 to 24 atoms per cell. Contrasting with other datasets used in~\cref{tab:gen}, where compositions typically correspond to a single stable structure, \textbf{Carbon-24} features a wide array of structures for any given composition. This dataset allows us to evaluate the capability to generate diverse one-to-many metastable structures, reflecting the variability inherent in crystal structures.

\textbf{Extending \model{} to Ab Initio Generation Task} We utilize the approach described in Appendix G of the DiffCSP literature~\cite{jiao2023crystal} to extend \model{} to the ab initio generation task.

\textbf{Baseline.} Our approach is compared against four generative methods suited to this dataset. \textbf{FTCP}\citep{REN2021}, a coordinate-based, non-E(3)-invariant method, represents crystals via a blend of real-space and Fourier-transformed properties, utilizing a CNN-VAE architecture for generation. \textbf{G-SchNet}\citep{NIPS2019_8974} employs an autoregressive model for structure generation, while \textbf{P-G-SchNet} is a G-SchNet variant incorporating periodicity. \textbf{CDVAE}\citep{xie2021crystal}, as previously mentioned, integrates a score matching-based decoder into the VAE framework; here, its standard version is applied without modifications. 
\textbf{SyMat}~\cite{luo2023towards} uses a variational auto-encoder for generating periodic structures, defining lattice and atom types. \textbf{DiffCSP}~\cite{jiao2023crystal}, a diffusion method, learns stable structure distributions, incorporating translation, rotation, and periodicity, effectively modeling material systems.

\textbf{Evaluation Metrics} We assess the results using three different criteria. \textbf{Validity}: This encompasses both structural and compositional validity. Structural validity is assessed by calculating the percentage of generated structures where all pairwise distances exceed 0.5 Å, while compositional validity checks for charge neutrality using the SMACT criteria~\cite{davies2019smact}. \textbf{Coverage}: This metric evaluates how well the structural and compositional attributes of the generated samples $\gS_g$ match those in the test set $\gS_t$. It uses $d_S(\gM_1,\gM_2)$ and $d_C(\gM_1,\gM_2)$ to represent the L2 distances for CrystalNN structural fingerprints~\citep{zimmermann2020local} and normalized Magpie compositional fingerprints~\citep{ward2016general}, respectively. Coverage Recall (COV-R) is calculated as $\text{COV-R}=\frac{1}{|\gS_t|}|\{\gM_i| \gM_i\in \gS_t, \exists \gM_j\in\gS_g, d_S(\gM_i,\gM_j)<\delta_S, d_C(\gM_i,\gM_j)<\delta_C\}|$, with predefined thresholds $\delta_S, \delta_C$. Coverage Precision (COV-P) is defined in a similar manner but with the roles of $\gS_g$ and $\gS_t$ reversed. \textbf{Property Statistics}: This includes the calculation of Wasserstein distances for three properties—density, formation energy, and elemental count—between the generated and test structures, denoted as $d_\rho$, $d_E$, and $d_{\text{elem}}$, respectively. The validity and coverage metrics are based on 10,000 generated samples, whereas the property statistics are derived from 1,000 samples that passed the validity check.

\textbf{Results.} Our method, \model{}, exhibits outstanding performance across multiple metrics, as detailed in Table~\ref{tab:gen}. Notably, \model{} achieves competitive results in validity and coverage precision, underscoring the high quality of the samples it generates. Additionally, it delivers robust coverage recall, demonstrating the diversity of the structures produced. In the realm of property metrics, \model{} excels by significantly reducing the density distance $d_\rho$, influenced by the volume of the generated lattice, and the formation energy distance $d_E$, which relates to the atomic configuration. These achievements in minimizing key distances underscore the effectiveness of our symmetry-aware processing approach.

\begin{table*}[!tp]
\vskip -0.1in
\caption{Results on ab initio generation task. The results of baseline methods are from Jiao~\cite{jiao2023crystal}}
\label{tab:gen}
    \small
   \setlength{\tabcolsep}{3.2mm}
\centering
\resizebox{0.9\linewidth}{!}{
\begin{tabular}{ll|ccccccc}
\toprule
\multirow{2}{*}{\bf Data} & \multirow{2}{*}{\bf Method}  & \multicolumn{2}{c}{\bf Validity (\%)  $\uparrow$}  & \multicolumn{2}{c}{\bf Coverage (\%) $\uparrow$}  & \multicolumn{3}{c}{\bf Property $\downarrow$}  \\
& & Struc. & Comp. & COV-R & COV-P & $d_\rho$ & $d_E$ & $d_{\text{elem}}$ \\
\midrule
\multirow[t]{6}{*}{Perov-5} & FTCP & 0.24 & 54.24 & 0.00 & 0.00 & 10.27 & 156.0 & 0.6297 \\
& Cond-DFC-VAE & 73.60 & 82.95 & 73.92 & 10.13 & 2.268 & 4.111 & 0.8373 \\
& G-SchNet  & 99.92 & 98.79 & 0.18 & 0.23 & 1.625 &	4.746 &	0.0368 \\
& P-G-SchNet & 79.63 & \textbf{99.13} & 0.37 & 0.25 & 0.2755 & 1.388 & 0.4552 \\
& CDVAE & \textbf{100.0} & 98.59 & 99.45 & 98.46 & 0.1258 & 0.0264 & 0.0628\\
& SyMat & \textbf{100.0} & 97.40 & 99.68 & 98.64 & 0.1893 & 0.2364 & 0.0177\\
& DiffCSP & \textbf{100.0} & 98.85 &	\textbf{99.74} &	98.27 & \textbf{0.1110} & 0.0263 & \textbf{0.0128} \\
& \model{} & \textbf{100.0} & 98.60 &	99.60 & \textbf{98.76} & \textbf{0.1110} & \textbf{0.0257} & 0.0503 \\
\midrule

\multirow[t]{5}{*}{Carbon-24} & FTCP  & 0.08 & -- & 0.00 & 0.00 & 5.206 & 19.05 & -- \\
& G-SchNet & 99.94	 & --	  & 0.00 & 0.00 & 0.9427 & 1.320 &	-- \\
& P-G-SchNet & 48.39 & -- & 0.00 & 0.00 & 1.533 & 134.7 & --  \\
& CDVAE & \textbf{100.0} & -- & 99.80 & 83.08 & 0.1407 & 0.2850 & -- \\
& SyMat & \textbf{100.0} & -- & 100.0 & \textbf{97.59} & 0.1195 & 3.9576 & -\\
& DiffCSP & \textbf{100.0} & -- & \textbf{99.90} & 97.27 & 0.0805 & 0.0820 & -- \\
& \model{} & \textbf{100.0} & -- &	99.75 & 97.12 & \textbf{0.0734} & \textbf{0.0508} & -- \\

\midrule

\multirow[t]{5}{*}{MP-20} & FTCP & 1.55 & 48.37 & 4.72 & 0.09 & 23.71 & 160.9 & 0.7363 \\
& G-SchNet & 99.65 &	75.96 & 38.33 & 99.57 &	3.034 &	42.09 &	0.6411	\\
& P-G-SchNet & 77.51 & 76.40 & 41.93 & 99.74 & 4.04 & 2.448 & 0.6234 \\
& CDVA & \textbf{100.0} & 86.70 & 99.15 & 99.49 & 0.6875 & 0.2778 & 1.432  \\
& SyMat & \textbf{100.0} & \textbf{88.26} & 98.97 & \textbf{99.97} & 0.3805 & 0.3506 & 0.5067\\
& DiffCSP & \textbf{100.0} & 83.25 & \textbf{99.71} & 99.76 & 0.3502 & 0.1247 & \textbf{0.3398} \\
& \model{} & 99.97 & 82.20 & 99.65 & 99.68 & \textbf{0.1300} & \textbf{0.0848} & 0.3978 \\

\bottomrule
\end{tabular}
}

\vskip -0.2in
\end{table*}



\section{Visualizations}
\begin{figure}[h]
    \centering    \includegraphics[width=0.9\textwidth]{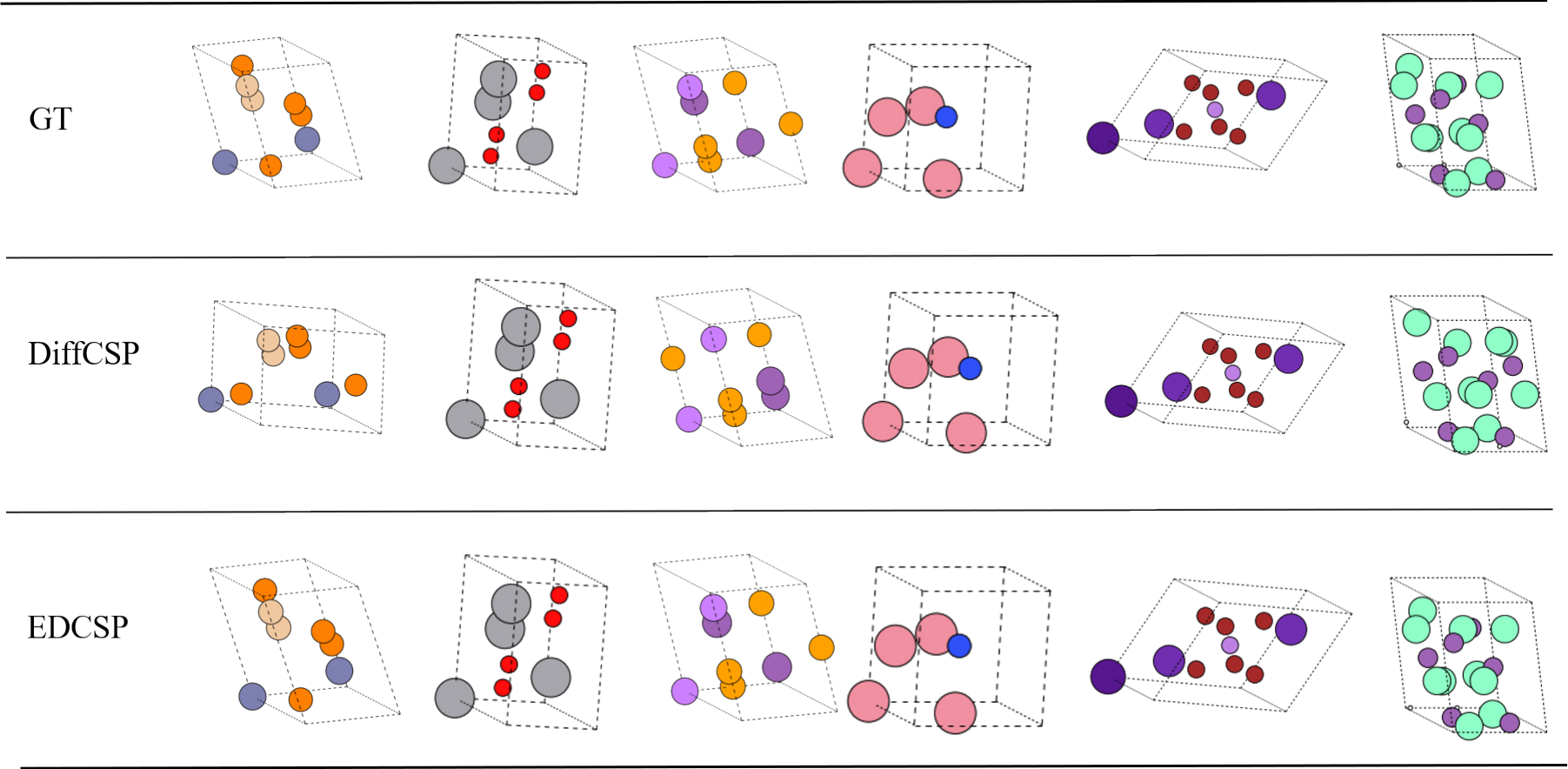}
    \caption{Additional visualizations of the predicted structures. We translate the same atom to the origin for better visualization and comparison.}
    \label{fig:vis_apd}
\end{figure}

In this section, we present additional visualizations of the predicted structures from \model{} and the second best method DiffCSP in Figure~\ref{fig:vis_apd}. Our \model{} provides more accurate predictions compared with DiffCSP.

